\documentclass[ floatfix,reprint,amsmath,amssymb,aps,prb]{revtex4-1}
\usepackage{xcolor}
\usepackage{bm}
\usepackage{textcomp}
\usepackage{amsmath}
\usepackage{amssymb}
\usepackage{graphicx}
\usepackage{esint}
\usepackage{color}
\usepackage{float}
\usepackage{enumitem}

\begin{document}
\title{Rigorous Wave-optical Treatment of Photon Recycling in Thermodynamics of Photovoltaics: The Case of Perovskite Thin-Film Solar Cells
}

\author{Muluneh G. Abebe $^1$, Aimi Abass$^{1,*}$,Guillaume Gomard$^{2,3}$,  Lin Zschiedrich$^{4,5}$, Uli Lemmer$^{2,3}$, Bryce S. Richards$^{2,3}$, Carsten Rockstuhl$^{1,6}$, and Ulrich W. Paetzold$^{2,3}$}
\affiliation{\mbox{$^1$Institute of Nanotechnology, Karlsruhe Institute of Technology (KIT), 76021 Karlsruhe, Germany} 
\mbox{$^2$Light Technology Institute, Karlsruhe Institute of Technology (KIT), 76128 Karlsruhe, Germany}
\mbox{$^3$Institute of Microstructure Technology, Karlsruhe Institute of Technology (KIT), 76021 Karlsruhe, Germany} 
\mbox{$^4$JCMwave GmbH, Bolivarallee 22, D – 14 050 Berlin, Germany }
\mbox{$^5$Zuse Institute Berlin, Takustrasse 7, 14195 Berlin, Germany}
\mbox{$^6$Institute of Theoretical Solid-State Physics, Karlsruhe Institute of Technology (KIT), 76128 Karlsruhe, Germany}
$^*$Corresponding author: aimi.abass@kit.edu}

\begin{abstract}

The establishment of a rigorous theory on thermodynamics of light management in photovoltaics that accommodates various loss mechanisms as well as wave-optical effects in the absorption and reemission of light is at stake in this contribution. To this end, we propose a theoretical framework to calculate the open-circuit voltage enhancement resulting from photon recycling ($\Delta V^{\mathrm{PR}}_{\mathrm{oc}}$) with rigorous wave-optical treatment. It can be applied to both planar thin-film and nanostructured single-junction solar cells. We derive an explicit expression for $\Delta V^{\mathrm{PR}}_{\mathrm{oc}}$, which reveals its dependence on internal quantum luminescence efficiency, parasitic reabsorption, and on photon escape probabilities of reemmited photons. While the internal quantum luminescence efficiency is an intrinsic material property, both latter quantities can be determined rigorously for an arbitrary solar cell architecture by three-dimensional electrodynamic dipole emission calculations.  We demonstrate the strengths and validity of our framework by determining the impact of photon recycling on the $V_{\mathrm{oc}}$ of a conventional planar organo-metal halide perovskite thin-film solar cell and compare it to established reference cases with perfect antireflection and Lambertian light scattering.  Our calculations reveal $\Delta V^{\mathrm{PR}}_{\mathrm{oc}}$ values of up to 80 mV for the considered device stack in the absence of angular restriction and up to 240 mV when the escape cone above the cell is restricted to $\theta_{\mathrm{out}}=2.5^\circ$ around the cell normal. These improvements impose severe constraints on the parasitic absorption as a parasitic reabsorption probability of only 2\% reduces the $\Delta V^{\mathrm{PR}}_{\mathrm{oc}}$ to 100 mV for the same angular restriction. Our work here can be used to provide design guidelines.

\end{abstract}
\maketitle

\section{Introduction}
Photon recycling in solar cells refers to charge carrier generation in the photovoltaic (PV) active material by reabsorption of photons that originate from radiative recombination within the semiconductor \cite{Parrott1993, Badescu1997, Zeng2016,   Kirchatz2016}. Although radiative recombination itself is inherently present in all PV materials, being the reversible process of light absorption and charge carrier generation \cite{Roosbroeck1954,Dumke1957}, photon recycling is only relevant to solar cells employing absorber materials with very low non-radiative recombination losses and high internal quantum luminescence efficiencies ($Q_{\mathrm{i}}^{\mathrm{lum}}$) \cite{Roosbroeck1954, Badescu1997, Smestad1992, Parrott1993, Green2012}. Next to this, efficient photon recycling in solar cells requires significant reabsorption of radiatively emitted photons, which implies low parasitic absorption losses and  light confinement (see Fig. 1). The latter is altered as soon as light trapping structures are integrated into the PV device that trap the incident light but also enable the reverse process, the outcoupling of photons generated by radiative recombination \cite{Miller2012, Rau2014, Staub2017}. Due to these strict requirements, for the case of unconcentrated solar irradiation, only planar solar cells based on epitaxially grown crystalline semiconductors such as GaAs were expected to significantly benefit from an enhanced performance due to photon recycling \cite{Steiner2013, Braun2013, Kosten2014,Vossier2015}. For a planar GaAs solar cell stack without any angular restriction, Walker \textit{et al.} showed that ignoring photon recycling may lead to an underestimation of the $V_{\mathrm{oc}}$ by 1.9\% of its value \cite{Walker2015}. By utilizing a narrow band dielectric multilayer angular restrictor, Kosten \textit{et al.} experimentally measured a $V_{\mathrm{oc}}$ enhancement due to enhanced photon recycling in a GaAs solar cell reaching 3.6 mV \cite{Kosten2014}.
\begin{figure}[h!] \label{Fig1}
	\includegraphics[scale=0.45]{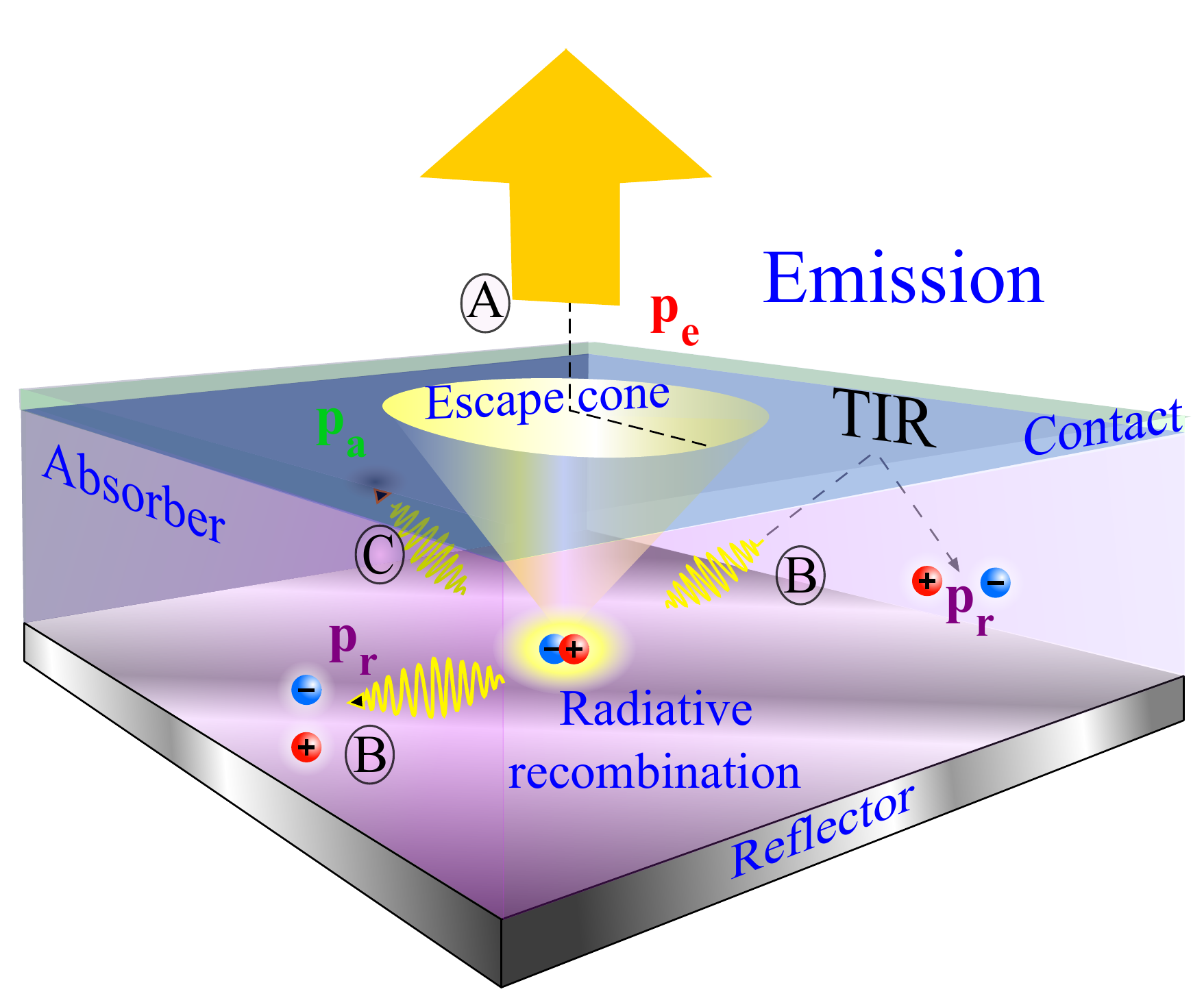}
	\caption{Possible routes that reemitted photons may take following radiative recombination inside the active material: In case A, a radiatively emitted photon propagates within the escape cone, possibly after multiple passes through the active layer, and escapes into free space without being recycled. In case B, photon recycling occurs due to  a possible direct reabsorbtion or upon total internal reflection (TIR) for photons emitted outside escape cone. Finally, in case C, a reemitted photon travels in the semiconductor for some distance before it gets parasitically absorbed in another layer without generating charge carriers. The probabilities that case A, B or C occurs are denoted by $p_\mathrm{e}$, $p_\mathrm{r}$, and $p_\mathrm{a}$ respectively. The probability quantities discussed here fulfills the relation  $p_\mathrm{r}+p_\mathrm{a}+p_\mathrm{e}=1$.}
    \end{figure}
    
However, with the recent fast rise of organo-metal halide perovskite solar cells, another highly efficient solution-processable multicrystalline PV material that could benefit from photon recycling emerged \cite{Saliba2015,Kirchatz2016}. Less than 5 years after the first reported solid state organo-metal halide perovskite solar cells, a record power conversion efficiency of 22.7\% was reported\cite{NREL}. The promise of perovskite solar cells is founded in their close to optimal combination of optical and electrical material properties, combining high absorption coefficients and long diffusion lengths \cite{Stranks2013, DeWolf2014, QIU2015, Joong2015,Yang2015,Saliba2016,Peng2017}. Very high radiative recombination rates and low non-radiative rates, approaching those measured in GaAs absorbers, were evidenced in organo-metal halide perovskites such as methylammonium lead triiodide (CH$_3$NH$_3$PbI$_3$) \cite{deQuilettes2016,Crothers2017,Davies2018}. According to first estimations, the $Q_{\mathrm{i}}^{\mathrm{lum}}$ of organo-metal halide pervoskites can surpass at least 70\% at one sun of solar irradiation and reaching even larger $Q_{\mathrm{i}}^{\mathrm{lum}}$ with stronger irradiation \cite{ Richter2016}. Moreover, in thin-film perovskite solar cells, radiatively emitted photons are reported to propagate over distances of few tens of microns experiencing multiple reabsorption and reemission events, which is an order of magnitude longer than the device thickness \cite{Pazos2016}. 

The high $Q_{\mathrm{i}}^{\mathrm{lum}}$ in perovskite PV results in an increasing interest on photon recycling in such solar cells \cite{Kirchatz2016, Tress2017}.
In a recent work, Kirchartz \textit{\textit{et al.}} predicted an approximate maximum possible $\Delta V^{\mathrm{PR}}_{\mathrm{oc}}$ in the radiative limit up to around 50-100 mV for devices with poor outcoupling efficiencies (e.g. planar devices) \cite{Kirchatz2016}. For devices with efficient outcoupling (for example devices with anti-reflection coatings and light trapping textures), Kirchartz \textit{\textit{et al.}} predicted a $\Delta V^{\mathrm{PR}}_{\mathrm{oc}}$ between 10-40 mV. While in their calculations they treated light absorption rigorously, the parasitic reabsorption probability remained a parameter and the emission probability was approximated  assuming an angle independent absorption responses and studying the system in the framework of ray optics. However, for thin-film planar stacks as well as nanostructures solar cells, coherent effects need to be considered. This immediately prompts for a rigorous treatment in the wave optics regime, thus solutions to Maxwell's equations need to be considered.

In order to render the rigorous treatment of photon recycling in arbitrary device architecture possible, we establish in this contribution a rigorous theory on the thermodynamics of light management in PV. We derive an explicit expression for the open-circuit voltage enhancement due to photon recycling ( $\Delta V^{\mathrm{PR}}_{\mathrm{oc}}$) under realistic conditions. The derived expression describes the dependence of $\Delta V^{\mathrm{PR}}_{\mathrm{oc}}$ on the intrinsic material property $Q_{\mathrm{i}}^{\mathrm{lum}}$, the parasitic reabsorption probability of radiatively emitted photons ($p_\mathrm{a}$), and the escape probability of radiatively emitted photons ($p_\mathrm{e}$). These device architecture specific probabilities are typically calculated with a simplified ray optical model or introduced phenomenologically in current available analysis in literature \cite{Rau2014,Walker2015,Kirchatz2016,Kirchartz2018}. Here, however, we do not just introduce these probabilities on phenomenological grounds but determine them through rigorous numerical three-dimensional electromagnetic dipole emission calculations. We therefore provide a framework for treating the impact of photon recycling on $ V_{\mathrm{oc}}$ rigorously. 

We demonstrate the application and validity of our theoretical framework by determining the impact of photon recycling on the open-circuit voltage $V_{\mathrm{oc}}$ of organo-metal halide perovskite thin-film solar cells and compare it to established reference cases with perfect antireflection and Lambertian light scattering. In order to discriminate the impact of various realistic physical effects on the $\Delta V^{\mathrm{PR}}_{\mathrm{oc}}$, we present stepwise analyses with increasing complexities, considering rigorously imperfect absorption, parasitic reabsorption, angular dependence, and reemitted photon escape probability. 

\section{Theory}

This work builds upon the comprehensive theory of thermodynamics of light management in photovoltaics developed by Rau \textit{et al.} \cite{Rau2014}. Their theoretical analysis includes a clear discrimination of the various entropic loss processes that reduce the $V_{\mathrm{oc}}$ for realistic single-junction solar cells experiencing parasitic absorption, incomplete light absorption, and non-radiative recombination. We extend their formalism to a framework that renders the treatment of angle dependent absorption responses possible and provides the opportunity to exploit input from rigorous dipole emission calculations to obtain a more accurate prediction of the effect of photon recycling.

\subsection{The open-circuit voltage}
In this section, we review the fundamentals of the $V_{\mathrm{oc}}$ in the framework of the detailed balance (DB) theory, which serves as a starting point for the deduction of the impact of photon recycling.

First, we inspect the ideal case in the absence of non-radiative recombination and parasitic reabsorption, which is given by the Shockley-Queisser limit  and describes the $V_{\mathrm{oc}}$ in the radiative limit ($V^{\mathrm{rad}}_{\mathrm{oc}}$) \cite{SQ1961}. The open-circuit condition is met when the recombination current density  is equal to the short-circuit (photo-generation) current density $J_{\mathrm{sun}}$. Thus, the current balance at open circuit bias reads as 
\begin{equation}\label{eq_currentbalance}
 J_{\mathrm{sun}} - J_{0,\mathrm{rad}}  \mathrm{exp}\lbrace(q V^{\mathrm{rad}}_{\mathrm{oc}})/(kT_{c}) \rbrace =0  
\end{equation}
where $J_{0,\mathrm{rad}}$ is the radiative dark saturation current, $k$ is the Boltzmann constant, $q$ is the electric charge of an electron,  and $T_{c}$ is the cell temperature. Note that Eq.~\ref{eq_currentbalance} inherently assumes Boltzmann statistics for the carrier densities in the absorber and thus is limited to solar cell architectures with electronically homogeneous absorbers describable by semi-classical bulk semiconductor physics.  Assuming perfect carrier collection, the short-circuit current density $J_{\mathrm{sun}}$ is given by 
\begin{equation}\label{eq_SCcurrent}
J_{\mathrm{sun}}= 2\pi q\int_{E_{\mathrm{g}}}^{\infty}\int_{0}^{\theta_{\mathrm{in}}} A(E,\theta)\phi_{\mathrm{sun}}(E)\sin(\theta)\cos(\theta)d\theta dE ,
\end{equation}
where $A(E,\theta)$ is the energy and angle dependent absorptance, $E_{\mathrm{g}}$ is the bandgap of the absorber material, $E$ is the photon energy, $\theta_{\mathrm{in}}$ is the incoming angle cone of the solar irradiation, and $\phi_{\mathrm{sun}}(E)$ is the incoming solar spectral photon flux, which corresponds to the number of incident photons from the sun per unit projected area, time, and solid angle $\Omega$. Here, we consider an absorption response that does not depend on the azimuthal angle. For the case of direct solar illumination, typically $\theta_{\mathrm{in}} =0.266^{\mathrm{o}}$ \cite{Wurfel2007}. 

At thermal equilibrium and in the absence of nonradiative recombination processes, $J_{0,\mathrm{rad}}$ is equal to the product of elementary charge and the photon flux emitted by the solar cell $J_{\mathrm{em}}$. Assuming that the solar cell emits as a blackbody and considering a time symmetric system, where emissivity can be considered to be equal to absorptivity regardless of  the spectral dependence \cite{Kirchoff1860,Tiedje1984,Kirchartz2008}, $J_{\mathrm{em}}$ can be written as
    \begin{equation}\label{eq_Emcurrent}
J_{\mathrm{em}}= 2\pi q\int_{E_{\mathrm{g}}}^{\infty}\int_{0}^{\theta_{\mathrm{out}}} A(E,\theta) n_{\mathrm{a}}^2 \phi_{\mathrm{bb}}(E)\sin(\theta)\cos(\theta)d\theta dE,	
\end{equation}
where $\phi_{\mathrm{bb}}$ is the black body spectral emission profile at the solar cell operating temperature, $n_{\mathrm{a}}$ is the refractive index of the ambient media, and $\theta_{\mathrm{out}}$ is the angle relative to the surface normal, which defines an escape cone within which photons can leave the solar cell. In this contribution, without limiting the general validity of our approach, we consider the case of air as ambient on the top (cladding refractive index $n_{\mathrm{a}}=1$). Thus, by rearranging Eq. 1 and utilizing $J_{0,\mathrm{rad}} = J_{\mathrm{em}}$, one obtains an expression for $V^{\mathrm{rad}}_{\mathrm{oc}}$ in the form 
\begin{equation}\label{eq_VocSQ}
 qV^{\mathrm{rad}}_{\mathrm{oc}}=kT_{c}\ln\left\lbrace \frac{J_{\mathrm{sun}} } {J_{\mathrm{em}}} \right\rbrace.
\end{equation}
  
\begin{figure}[h]
	\includegraphics[scale=0.3]{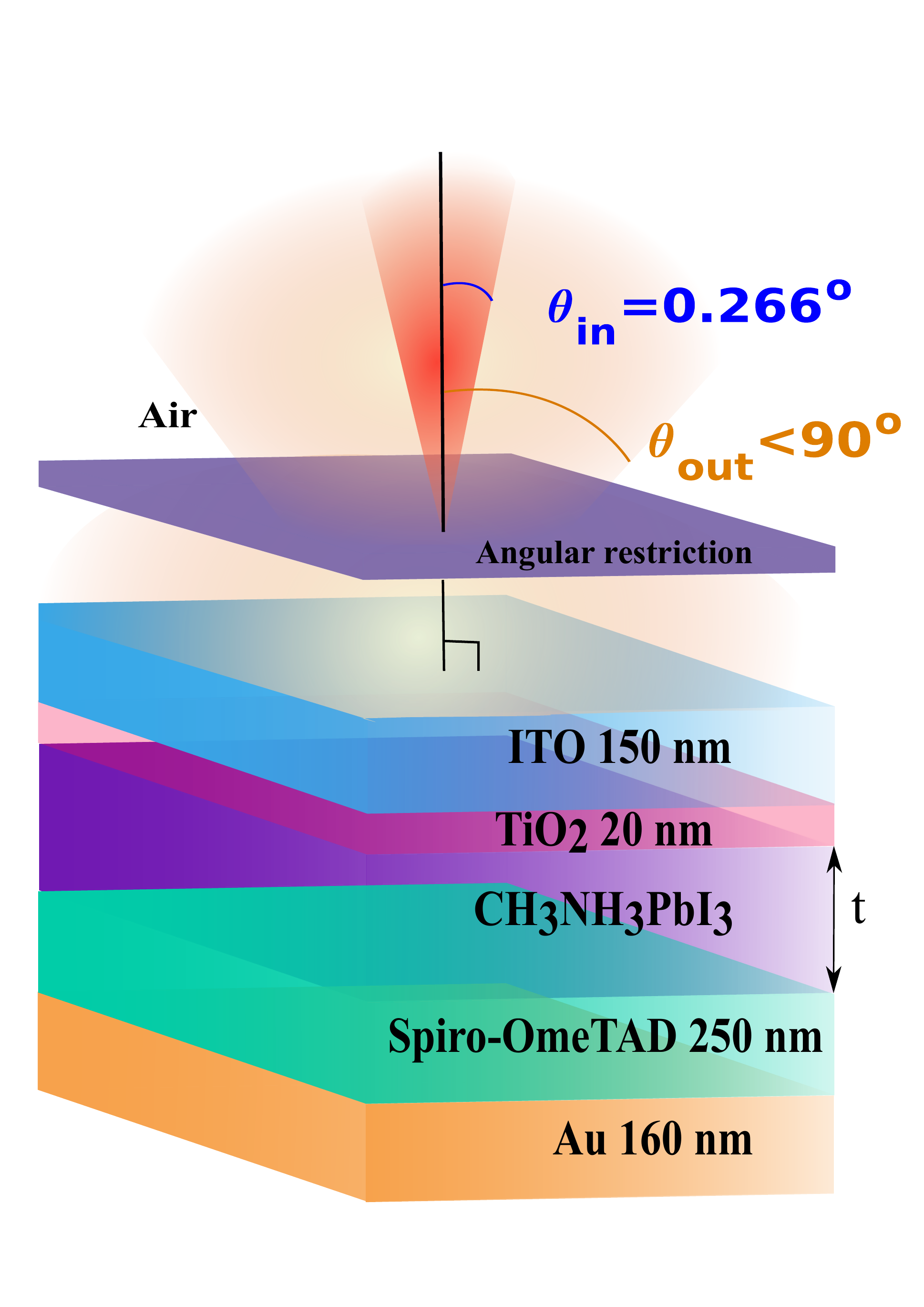}
	\caption{Schematic illustration of a layer stack organo-metal halide perovskite solar cell considered in this work. $\theta_{\mathrm{in}}$ is the incoming angle cone of the solar irradiation and $\theta_{\mathrm{out}}$ is the angle relative to the surface normal, which defines an escape cone within which photons can leave the solar cell. 
 \label{Fig2}}    
\end{figure}

 As shown in Fig.~\ref{Fig2}, the surface emits light confined to a certain solid angle element $d\Omega$ at an angle $\theta$ relative to its normal. 
	Second, in considering a realistic device where effects such as non-radiative recombination and parasitic absorption are encountered, the open-circuit voltage is given by\cite{Rau2007}
    	\begin{equation}\label{eq_VocQLED}
qV^{\mathrm{DB}}_{\mathrm{oc}}=qV_{\mathrm{oc}}^{\mathrm{rad}}+kT_{c}{\ln\left\lbrace Q_{\mathrm{e}}^{\mathrm{LED}}\right\rbrace}.
	\end{equation}
    The superscript DB indicates that detailed balance is assumed in considering $V_{\mathrm{oc}}^{\mathrm{rad}}$, and $Q_{\mathrm{e}}^{\mathrm{LED}}$ is the external quantum luminescence efficiency. In discerning the impact of photon recycling on $V_{\mathrm{oc}}$, it is useful to separate the internal parameters that determine  $Q_{\mathrm{e}}^{\mathrm{LED}}$. This can be achieved with more ease if one reformulates Eq.~\ref{eq_VocQLED} in terms of current densities and probability quantities ( as detailed in Appendix A) in the form  
    \begin{equation}\label{eq_VocDB}
{qV^{\mathrm{DB}}_{\mathrm{oc}}=kT_{c}\ln{\left\lbrace\frac{J_{\mathrm{sun}}}{J_{\mathrm{em}}+J_{\mathrm{re,rad}}\left[(\frac{1}{Q_{\mathrm{i}}^{\mathrm{lum}}}-1)+p_{\mathrm{a}} \right]}\right\rbrace}}.
	\end{equation}\\ 
Equation~\ref{eq_VocDB} is a general expression for the open-circuit voltage of a single-junction cellin terms of the current densities. The accuracy depends on the way by which the latter are determined provided that current densities are calculated rigorously. An arbitrary absorption response can be considered, if the integrals in Eqs. \ref{eq_SCcurrent}-\ref{eq_Emcurrent} are determined numerically. For an optically homogeneous absorber, the total radiative recombination current within the absorber volume ($J_{\mathrm{re,rad}}$) can be described with \cite{Roosbroeck1954}
\begin{equation}\label{eq_Totradrecombcurrent}
J_{\mathrm{re,rad}}=  qt4\pi n^{2}\int_{E_{\mathrm{g}}}^{\infty}\alpha(E)\phi_{\mathrm{bb}}(E) dE ,	
\end{equation} 
where $t$ is the thickness of the solar cell absorber, $n$ is the real part of the refractive index of the absorber, and $\alpha$ is the extinction coefficient of the absorber medium. A more rigorous treatment can be done by considering a spatial integration of the local radiative rate as also done in this work. In Section III, we will consider relevant idealized and extreme absorption cases that are useful for exploring the limiting conditions of a solar cell's performance.

It shall be noted that Eqs.~\ref{eq_VocSQ}-\ref{eq_VocDB} already include the contribution of photon recycling if $n_\mathrm{a}$ is smaller than the refractive index of the absorber. Considering $n_\mathrm{a}$ with values larger than the absorber refractive index can lead to unphysical conditions in which $J_{\mathrm{em}} > J_{\mathrm{re,rad}}$. In Eq.~\ref{eq_VocSQ}, the impact of photon recycling is contained within the fact that $J_{\mathrm{em}} \leq J_{\mathrm{re,rad}}$. Under non-ideal conditions described by Eq.~\ref{eq_VocQLED}, the impact of photon recycling is additionally affected by $Q_{\mathrm{e}}^{\mathrm{LED}}$. 
    
\subsection{Photon recycling under realistic conditions}

The impact of photon recycling on open-circuit voltage ($\Delta V^{\mathrm{PR}}_{\mathrm{oc}}$) has been approximated in previous contributions such as by Kirchartz \textit{et al.} \cite{Kirchatz2016},  by a non-rigorous treatment of $p_\mathrm{a}$ and $p_\mathrm{e}$. In this work, we do not make such approximations but rely on rigorous calculation of  $\Delta V^{\mathrm{PR}}_{\mathrm{oc}}$ in arbitrary single-junction solar cells in the presence of non-radiative recombination and  parasitic photon reabsorption.

Therefore, we first note that the total bulk recombination current density $J_{\mathrm{re}}$ may have both radiative ($J_{\mathrm{re,rad}}$) and non-radiative components ($J_{\mathrm{re,nrd}}$) such that it has to be written as
\begin{equation}\label{eq_ra2}
J_{\mathrm{re}}=J_{\mathrm{re,rad}}+J_{\mathrm{re,nrd}}.
\end{equation}  
$J_{\mathrm{re,nrd}}$ represents here the sum of all possible non-radiative recombination current densities, given by trap assisted Shockley-Read-Hall and Auger recombination  losses\cite{Sze2006}. Rather than considering the detailed semiconductor aspects of non-radiative losses, we directly consider the intrinsic material property $Q_{\mathrm{i}}^{\mathrm{lum}}$, which can be deduced for a particular absorber material from photoluminescence measurements. This pragmatic consideration stems from the difficulty of retrieving exact parameter values from  solving non-linear coupled  semiconductor equations. Following the argument in Appendix A for Eq.~\ref{Rnrd}, $J_{\mathrm{re,nrd}}$ can be written as
\begin{equation}\label{eq_ra3}
J_{\mathrm{re,nrd}}=\frac{(1-Q_{\mathrm{i}}^{\mathrm{lum}})  qt 4\pi n^2\int_{E_{\mathrm{g}}}^{\infty} \alpha(E)\phi_{\mathrm{bb}}(E)dE}{Q_{\mathrm{i}}^{\mathrm{lum}}}. 
\end{equation}
Utilizing Eq.~\ref{eq_Totradrecombcurrent} and Eq.~\ref{eq_ra3}, Eq.~\ref{eq_ra2} can be expressed as
\begin{equation}\label{eq_ra4}
J_{\mathrm{re}}=\frac{qt 4\pi n^2 \int_{E_{\mathrm{g}}}^{\infty} \alpha(E)\phi_{\mathrm{bb}}(E)dE}{Q_{\mathrm{i}}^{\mathrm{lum}}}.
\end{equation} \\
Keeping in mind the total saturated current density as described by Eq.~\ref{eq_ra4} and making use of Eq.~\ref{eq_SCcurrent}, the open-circuit voltage of a solar cell in the absence of photon recycling $V^{\mathrm{SD}}_{\mathrm{oc}}$ can be calculated using the standard diode (SD) model \cite{Green1982B} with the expression
\begin{equation}\label{eq_VOCSD}
 qV^{\mathrm{SD}}_{\mathrm{oc}}=kT_{c}\ln\left\lbrace \frac{J_{\mathrm{sun}} } {J_{\mathrm{re}}} \right\rbrace.
\end{equation}
Equation \ref{eq_VOCSD} essentially assumes that the whole radiative portion of the recombination within the volume of the absorber is counted as loss. Therefore, $\Delta V^{\mathrm{PR}}_{\mathrm{oc}}$ can be deduced by using $\Delta V^{\mathrm{PR}}_{\mathrm{oc}} = V^{\mathrm{DB}}_{\mathrm{oc}}-V^{\mathrm{SD}}_{\mathrm{oc}}
$. We thus arrive at the key equation of the paper, which is the expression
\begin{equation}\label{eq_DVOCPR}
\boxed{q\Delta V^{\mathrm{PR}}_{\mathrm{oc}}  =kT_{c}{\ln\left\lbrace\frac{1 }{1-(1-p_{\mathrm{e}}-p_{\mathrm{a}})Q_{\mathrm{i}}^{\mathrm{lum}} }\right\rbrace}}
\end{equation}
where we utilize the fact that $p_{\mathrm{e}}$ can also be deduced by taking the ratio of the number of photons leaving the cell and what is generated inside the absorber volume and thus 
\begin{equation}\label{eq_pegen}
p_{\mathrm{e}}=J_{\mathrm{em}}/ J_{\mathrm{re,rad}}. 
\end{equation}

\subsection{Rigorous calculation of parasitic reabsorption and escape probabilities of reemitted photons}

In the literature, the probability quantities $p_{\mathrm{e}}$ and $p_{\mathrm{a}}$ are either considered as a phenomenological parameter or deduced assuming a black-body emission with no angular dependency \cite{Rau2014,Kirchatz2016}. We can extend this when evaluating $p_{\mathrm{e}}$ by considering an angular dependent absorption response in evaluating Eq.~\ref{eq_Emcurrent}. Upon inserting Eqs.~\ref{eq_Emcurrent} and ~\ref{eq_Totradrecombcurrent} to Eq.~\ref{eq_pegen}, one obtains the simplified expression for $p_{\mathrm{e}}$ based on black-body and DB considerations in the form
\begin{equation}\label{eq_pebb}
p_{\mathrm{e,bb}}= \frac{ \int_{E_{\mathrm{g}}}^{\infty}\int_{0}^{\theta_{\mathrm{out}}} A(E,\theta) n_{\mathrm{a}}^2 \phi_{\mathrm{bb}}(E)\sin(\theta)\cos(\theta)d\theta dE}
{t2n^{2}\int_{E_{\mathrm{g}}}^{\infty}\alpha(E)\phi_{\mathrm{bb}}(E) dE}.
\end{equation}

In order to deduce $p_{\mathrm{a}}$, one requires knowledge of the field distribution in the different layers, which typically requires a full-wave optical treatment. Due to this complexity, $p_{\mathrm{a}}$ is often considered as a parameter in previous thermodynamic analyses for the $V_\mathrm{oc}$. 

In this work, for the first time, we go a step further in accuracy by relying on rigorous dipole emission calculations in deducing the probability quantities.  More specifically, we deduce the total system Green's tensor $\underline{\boldsymbol{G}}(\boldsymbol{r},\boldsymbol{r}_0)$, which describes the optical response of the system due to a point source in our thin-film stack\cite{Paulus2000}. $\boldsymbol{r}_0$ is the position of the current dipole point source. Once the Green's tensor of the solar cell system is obtained, one can deduce the portion of the power leaving the device stack and absorption in the different layers and in turn deduce $p_{\mathrm{e}}$ and $p_{\mathrm{a}}$ rigorously. For example, $p_{\mathrm{e}}$ can be calculated through 
\begin{equation}\label{eq_averagepe}
p_{\mathrm{e,di}}= \left\langle  
\frac{\sum_o \int_{\mathrm{A}_\mathrm{out}} \boldsymbol{S}(\boldsymbol{r},\boldsymbol{r}_0)_{o}\cdot d\mathbf{A} } {\sum_o P^{\mathrm{dip}}_{o}(\boldsymbol{r}_0)}
\right\rangle_{\boldsymbol{r}_0}, 
\end{equation}
where $\boldsymbol{S}(\boldsymbol{r},\boldsymbol{r}_0)_{o}$ is the Poynting vector at position $\boldsymbol{r}$ due to a point dipole current source with polarization $o$ emitting at position $\boldsymbol{r}_0$. The index $o$ runs through all possible orientation of the dipole emitters. $\left\langle \right\rangle_{\boldsymbol{r}_0} $ describes an average across all positions from where emission can happen. $P^{\mathrm{dip}}_{o}$ is the total power emitted by a dipole given by\cite{Novotny12Book}
\begin{equation}\label{eq_DipolePower}
P^{\mathrm{dip}}_{o}(\boldsymbol{r}_0)=\frac{\omega |\boldsymbol{j}_o|^2}{2c^2 \epsilon} \left[ \boldsymbol{n}_o \cdot  \mathrm{Im} \left\lbrace  \underline{\boldsymbol{G}}(\boldsymbol{r}_0,\boldsymbol{r}_0) \right\rbrace \cdot \boldsymbol{n}_o \right],
\end{equation}
where $\epsilon$ is the permittivity and $\boldsymbol{n}_o$ is the unit vector of the current dipole polarization. $\boldsymbol{S}(\boldsymbol{r},\boldsymbol{r}_0)_{o}$ is given by
\begin{equation}\label{eq_PoyntingGreen}
\boldsymbol{S}(\boldsymbol{r},\boldsymbol{r}_0)_{o}=  \mathrm{Re}\left\lbrace i\omega \mu \underline{\boldsymbol{G}}(\boldsymbol{r},\boldsymbol{r}_0)\boldsymbol{j}_o(\boldsymbol{r}_0) \times \left[\nabla \times \underline{\boldsymbol{G}}(\boldsymbol{r},\boldsymbol{r}_0)\boldsymbol{j}_o(\boldsymbol{r}_0) \right] \right\rbrace,
\end{equation}
where $\mu$ is the magnetic permeability of the media and $\boldsymbol{j}_o(\boldsymbol{r}_0)$ is the current dipole point source oscillating at the angular frequency $\omega$.

The nominator in Eq.~\ref{eq_averagepe} describes the net power exiting the solar cell structure to the cladding material. The denominator gives the total power emitted by the considered dipole point source.  Thus, Eq.~\ref{eq_averagepe} refers to the relative portion of the emitted power which escapes the solar cell thereby giving the wave optical description of $p_{\mathrm{e}}$. Equation~\ref{eq_averagepe} inherently considers the impact of spatial dependent local density of photonic states, which is not done in Eq.~\ref{eq_pebb}. For a more rigorous wave-optical treatment, Eq.~\ref{eq_averagepe} must also take the weighted average over the spectral distribution of the emission. In practice, as the linewidth of the emission in our considered perovskite material is relatively narrow, we only evaluate Eq.~\ref{eq_averagepe} for the peak emission wavelength of 770~nm\cite{Staub2016}. Equation \ref{eq_averagepe} essentially represents the escaped power portion weighted against the emitted power averaged over all relevant position and orientation. An even more rigorous treatment, which takes into account the whole spatially-resolved information of the electronic occupation and optoelectronic coupling, can be done by incorporating the formalism presented by Aeberhard and Rau ~\cite{Aeberhard2017}. 

Similarly, $p_{\mathrm{a}}$ can be calculated rigorously by  
\begin{equation}\label{eq_averagepa}
p_{\mathrm{a}}= \left\langle  
\frac{\sum_o \sum_l \int_{V_l} \nabla \cdot \boldsymbol{S}(\boldsymbol{r},\boldsymbol{r}_0)_{o} d^3\boldsymbol{r} } {\sum_o P^{\mathrm{dip}}_{o}(\boldsymbol{r}_0)}
\right\rangle_{\boldsymbol{r}_0}, 
\end{equation}
The index $l$ runs through all layers in the device stack other than the absorber layer. $V_l$ indicates the volume of layer $l$. The terms within the integral in Eq.~\ref{eq_averagepa} is essentially the divergence of the Poynting vector which gives the net power loss per unit volume in the absence of gain in the system. The integration term in Eq.~\ref{eq_averagepa} thus gives the absorbed power in the supporting layer $l$ of the solar cell device stack.

It shall be noted that one can in principle readily consider possible spatial inhomogeneities of various electrical properties in the device stack by introducing a position dependent probability coefficient in calculating $p_{\mathrm{a}}$ and $p_{\mathrm{e}}$. While such additional complexities may be required in modelling the emission properties of certain nanostructured solar cells, this is typically not needed for planar multilayer systems.

Here, for proof of principle purposes, we will consider an examplary planar multilayer thin-film solar cell device stack. In such systems, the calculation of the total Green's tensor can be done analytically by working in Fourier space\cite{Paulus2000}. Further details of the rigorous dipole emission calculations we performed to obtain $p_{\mathrm{e}}$ and $p_{\mathrm{a}}$  are given in appendix C. To illustrate how wave-optical effects influence $p_{\mathrm{e}}$ and $p_{\mathrm{a}}$, we show in appendix C the dependence of $p_{\mathrm{e}}$ and $p_{\mathrm{a}}$ on the emitter's vertical position within the active layer.

\subsection{Internal quantum luminescence efficiency  requirements for efficient photon recycling}

Having presented the rigorous analysis of $\Delta V^{\mathrm{PR}}_{\mathrm{oc}}$, we discuss here the relevant regimes for photon recycling with respect to $Q_{\mathrm{i}}^{\mathrm{lum}}$. A proper understanding of these regimes is crucial in deciding whether one should care about photon recycling in the design of single-junction solar cells. We therefore proceed to examine the dependence of $V_{\mathrm{oc}}$  (Eq.~\ref{eq_VocDB})  and $\Delta V^{\mathrm{PR}}_{\mathrm{oc}}$ (Eq.~\ref{eq_DVOCPR}) on $Q_{\mathrm{i}}^{\mathrm{lum}}$ for an exemplary organo-metal-halide perovskite solar cell as shown in Fig. \ref{Fig2}. 
\begin{figure}[tbh]
	\includegraphics[scale=0.245]{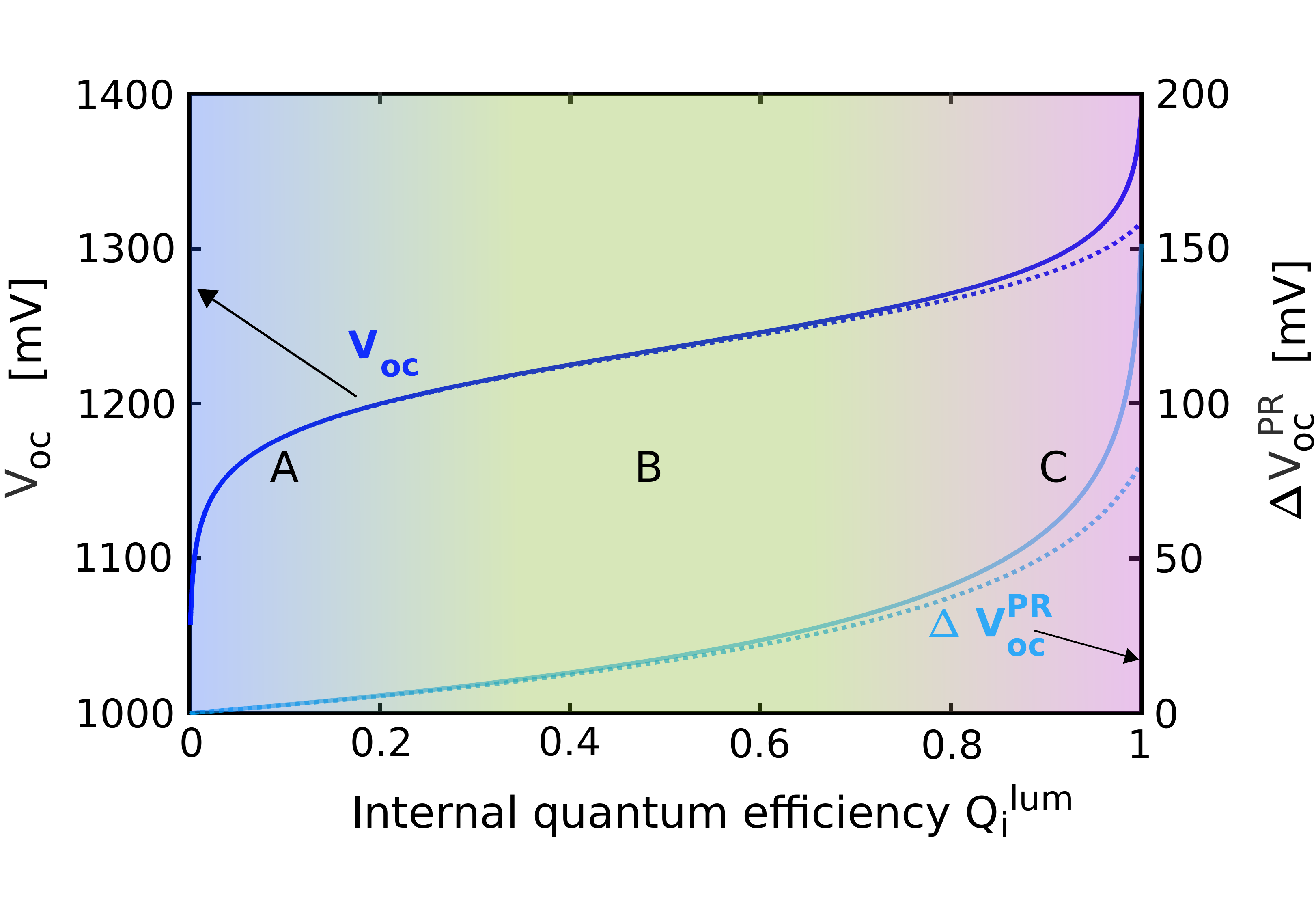}
	\caption
    { Three regions (here denoted A, B and C) can be distinguished when plotting $V_{\mathrm{oc}}$ as a function of $Q_{\mathrm{i}}^{\mathrm{lum}}$. The $V_{\mathrm{oc}}$ is calculated assuming no parasitic reabsorption ($p_{\mathrm{a}}=0$) for $\theta_{\mathrm{out}}= 90^\circ $ (dashed lines) and $\theta_{\mathrm{out}}= 15^\circ $ (solid lines).
 \label{Fig3}}    
\end{figure}

Three regimes are identified. Regime A describes the case of low $Q_{\mathrm{i}}^{\mathrm{lum}}$, where the impact of photon recycling is negligible. As Fig. \ref{Fig3} shows, $\Delta V^{\mathrm{PR}}_{\mathrm{oc}}$ vanishes as $Q_{\mathrm{i}}^{\mathrm{lum}}$ approaches zero. The superlinear reduction of the open-circuit voltage as $Q_{\mathrm{i}}^{\mathrm{lum}}$ is decreased below 0.1 is due to the strong non-radiative recombination. In regime B, photon recycling starts to have a significant impact on $V_{\mathrm{oc}}$, though mainly through single reemission and reabsorption events due to moderate values of $Q_{\mathrm{i}}^{\mathrm{lum}}$. In regime C, multiple reemission and reabsorption events are possible prior to either carrier extraction or reemitted photons escaping the device. As a consequence, a nonlinear increase of $V_{\mathrm{oc}}$ due to photon recycling occurs as $Q_{\mathrm{i}}^{\mathrm{lum}}$ approaches 1. This nonlinear increase is considerably stronger when the escape cone is reduced (here we compare $\theta_{\mathrm{out}}= 90^\circ $ and $\theta_{\mathrm{out}}= 15^\circ $). This makes regime C the most relevant for photon recycling. In the following sections, we will demonstrate that the strong dependence of the $V_{\mathrm{oc}}$ increase due to photon recycling in regime C will be heavily influenced by the photon escape and parasitic reabsorption probabilities upon reemission within the absorber, which can be additionally influenced through angular restriction.

\section{Results}

In the following, we apply the theoretical framework developed in the previous sections to introduce the rigorous treatment of the impact of photon recycling on $V_{\mathrm{oc}}$ for an exemplary organo-metal-halide perovskite thin-film solar cell [shown in Fig.~\ref{Fig2}]\cite{QIU2015,vanEerden2017}. For all the calculations discussed in this contribution, we utilized refractive index data obtained from measurements detailed in a recent publication \cite{vanEerden2017}. In order to discriminate the impact of various physical effects encountered in real-world PV devices on the photon recycling enhancement, we present analyses of the $\Delta V^{\mathrm{PR}}_{\mathrm{oc}}$ and $V_{\mathrm{oc}}$ in a stepwise increase of complexity considering:

\begin{enumerate}[label=\Alph*.]
\item \textit{Light trapping schemes} 
\item \textit{Accurate parasitic reabsorption probability} obtained from rigorous dipole calculations 
\item \textit{Angular dependence of the absorption response} 
\item \textit{Accurate escape probability} obtained from rigorous dipole calculations
\item \textit{Angular restriction} 
\end{enumerate}

In recent literature, the analysis of the impact of photon recycling on the $V_{\mathrm{oc}}$ has already considered the absorption response of various light trapping schemes (A) for an exemplary perovskite solar cell device architecture \cite{Kirchatz2016,Rau2014}. In these past contributions, Eq.~\ref{eq_pebb} was used assuming the absorption at different incoming angle to be the same as for normal incident. To some degree, this analysis has also considered parasitic reabsorption probabilities (step B), although they considered  $p_{\mathrm{a}}$ as a phenomenological parameter instead of deducing it from rigorous wave-optical calculations. For the sake of clarity, we first follow the foot steps of this contributions prior to going beyond their analysis with more rigorous treatments (steps C-E).

\subsection{Impact of light trapping schemes}

The starting point of the rigorous treatment of the impact of photon recycling on $V_{\mathrm{oc}}$ is the discussion of the impact of light trapping and absorption response of the perovskite absorber layer in a thin-film stack calculated with wave optics. We first examine $V_{\mathrm{oc}}$ (Eq.~\ref{eq_VocDB}) and $\Delta V^{\mathrm{PR}}_{\mathrm{oc}}$ (Eq.~\ref{eq_DVOCPR}) under the assumption of no parasitic reabsorption ($p_{\mathrm{a}}=0$), the absorber emits as a black-body in an angle independent manner, and without any angular filtering [results will be shown in Fig. \ref{Fig8}]. More specifically, we calculate $p_{\mathrm{e}}$ with Eq.~\ref{eq_pebb} while assuming $A(E,\theta)=A(E,0)$ without any angular restriction ($\theta_{\mathrm{out}}=90^\circ$ where emission to the top side full half space is allowed).   

\begin{figure}[h]
	\includegraphics[scale=0.45]{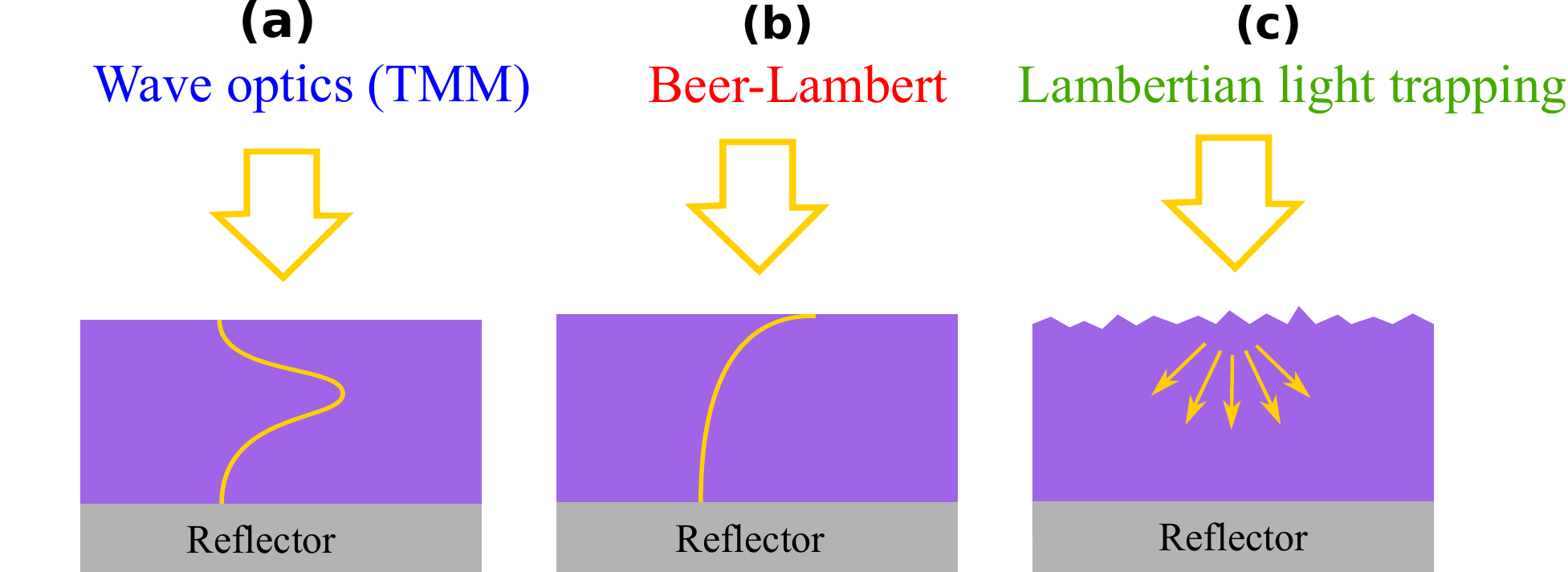}
	\caption{(a) Rigorous planar stack absorption response including interference deduced from rigorous calculations utilizing the transfer matrix method (TMM). (b) Beer-Lambert absorption response, which assumes zero front-side reflection. (c) Lambertian absorption response to consider optimum light trapping. \label{Fig4}}
\end{figure}

To provide a summary on the impact of different light trapping conditions, we analyze  $V_{\mathrm{oc}}$ and $\Delta V^{\mathrm{PR}}_{\mathrm{oc}}$  for three fundamental types of solar cell absorption responses (see Fig.~\ref{Fig4}). 
We consider first a planar multilayer experimental solar cell device stack depicted in Fig. \ref{Fig2} with absorptance retrieved from transfer matrix method (TMM) calculations [Fig. \ref{Fig4}a]\cite{BORN1980665}. We then compare the rigorously calculated absorption response to an ideal double-pass Beer-Lambert law assuming perfect incoupling of light, which is the ideal case of a perfect antireflection but no light trapping effects [Fig. \ref{Fig4}b]. Lastly, we consider a cell with a front-side random light trapping texture at the front of the cell [Fig. \ref{Fig4}c)]\cite{Tiedje1984,Green2012}. The scattering surface is considered to lead to a complete Lambertian randomization of light ray angles and results in an absorptance following the Yablonovich limit. The absorption spectra for each considered case at normal incidence for the case of a perovskite absorber with thickness $t=300$ nm  is given in Appendix B.

\begin{figure}
	\includegraphics[scale=0.24]{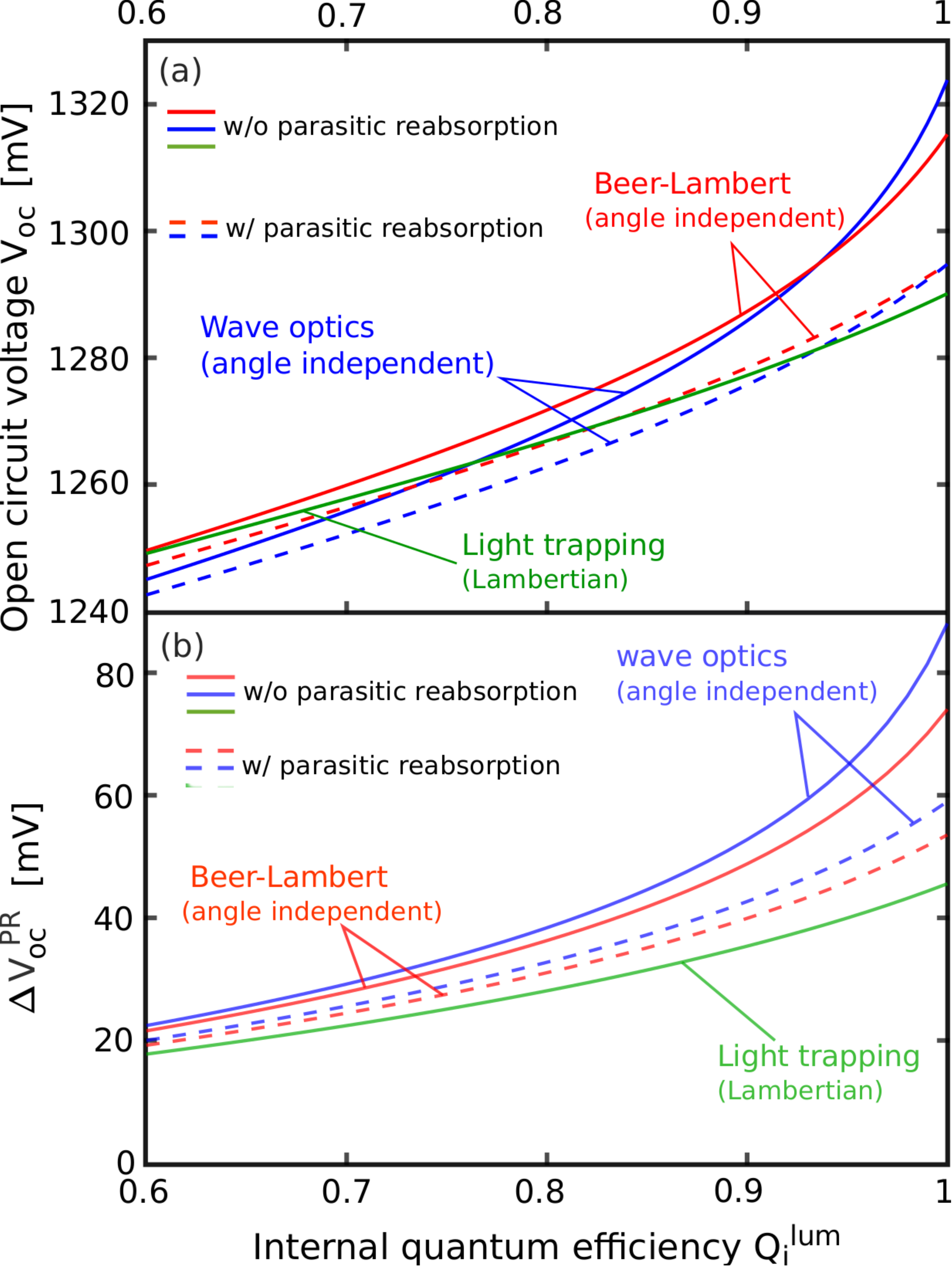}
	\caption{(a) Calculated $V_{\mathrm{oc}}$ and (b) $\Delta V^{\mathrm{PR}}_{\mathrm{oc}}$ as a function of $Q^{\mathrm{lum}}_{\mathrm{i}}$ for three different angle independent absorption responses: Beer-Lambert, TMM (wave optics), and Lambertian light trapping for $\theta_{\mathrm{out}}=\pi/2$ with and without parasitic reabsorption being present due to other layers in the solar cell. 
\label{Fig5}}
\end{figure}

The Lambertian response, which leads to the largest photon absorption (Appendix B), can be seen to provide the least open-circuit voltage ($V_{\mathrm{oc}}=1290$~mV at $Q_{\mathrm{i}}^{\mathrm{lum}}=1$) and $\Delta V^{\mathrm{PR}}_{\mathrm{oc}}$ (green line plots in Figs. \ref{Fig5}a and b). This is in line with time reversal considerations,  as strong absorption for a particular incoming direction translates to higher emission probability in that direction. The hypothetical architecture with a Lambertian texture allows for light incoming at all angles to be absorbed at the Lambertian limit. This omni-angle strong absorption translates to a system that supports high emission probability to all angles. This strong omni-angle emissivity translates to higher overall escape probability ($p^{\mathrm{L}}_{\mathrm{e},bb}=0.178$) for photons reemitted from within the absorber as compared to the other two cases ($p^{\mathrm{BL}}_{\mathrm{e},bb}=0.057$ for Beer-Lambert and $p^{\mathrm{TMM}}_{\mathrm{e},bb}=0.0332$ for wave optics). One additional reason why the architecture with the Lambertian scattering front texture has a larger $p^{\mathrm{L}}_{\mathrm{e},bb}$ is the loss of total internal reflection due to the texture.   

The $V_{\mathrm{oc}}$ deduced with full wave-optics consideration of the device stack (TMM absorption case, blue line plots in Fig. \ref{Fig5}) is lower as compared to the Beer-Lambert case (red line plots in Fig. \ref{Fig5}) for $Q_{\mathrm{i}}^{\mathrm{lum}}<0.95$. As $Q_{\mathrm{i}}^{\mathrm{lum}}$ approaches 1, however, the full wave-optics absorption case lead to a higher $V_{\mathrm{oc}}$ (up to $V_{\mathrm{oc}}=1324$~mv at $Q_{\mathrm{i}}^{\mathrm{lum}}=1$) as compared to the perfect-incoupling case (up to $V_{\mathrm{oc}}=1315$~mV). This is related to the fact that the organo-metal halide perovskite material has a high absorption coefficient over a major part of the spectral region above the bandgap. Having perfect light incoupling results in a much stronger sunlight absorption than the real full stack response, which in turn corresponds to a higher $J_{\mathrm{sun}}$ (Appendix B).  In turn, the perfect light incoupling assumption leads to a higher escape probability for reemitted photons compared to the wave optics response, thus to a smaller $\Delta V^{\mathrm{PR}}_{\mathrm{oc}}$ [Fig. \ref{Fig5}b]. This leads to the wave optics absorption case having a larger $V_{\mathrm{oc}}$ when $Q_{\mathrm{i}}^{\mathrm{lum}}>0.95$ where the impact of photon recycling is sufficiently large. 

The side conclusion that can be drawn here is that one must reduce $p_{\mathrm{e}}$, and hence the outcoupling efficiency, to really benefit from photon recycling. If the reduction of outcoupling efficiency is not followed by a reduction of short circuit current, which is possible if the absorption response within the acceptance cone for solar irradiation is left unchanged, one would obtain a higher open-circuit voltage.

We note in Fig. \ref{Fig5}(a) that the system with the Lambertian absorption response can actually possess a higher $V_{\mathrm{oc}}$  at lower $Q^{\mathrm{lum}}_{\mathrm{i}}$  where the impact of photon recycling is minimal.  This is due to the Lambertian response providing a larger short circuit current thus contributing to an increase in  $V_{\mathrm{oc}}$  as can be seen in Eq. 4.  Provided that the nonradiative recombination condition is identical for the different cells considered in Fig. 4, a solar cell with a Lambertian response can have a larger $V_{\mathrm{oc}}$  in the $Q^{\mathrm{lum}}_{\mathrm{i}}$ regime where photon recycling is not significant. 

\subsection{Accurate parasitic reabsorption obtained from rigorous dipole calculations }

The different light trapping schemes considered in subsection A essentially impact $p_\mathrm{e}$. In order to deduce the impact of photon recycling, one must also determine the value of $p_\mathrm{a}$ in the considered system, which we discuss in the following.  

From rigorous analytical dipole emission calculations\cite{Paulus2000} and utilizing Eq.~\ref{eq_averagepa}, we deduce $p_{\mathrm{a}}=0.069$ for the layer stack shown in Fig. \ref{Fig2} assuming an absorber thickness of $t=300$~nm. A major fraction of the parasitic reabsorption occurs in the indium tin oxide (ITO) layer. Though the realistic $p_{\mathrm{a}}$ value is relatively small, it causes a significant reduction of $\Delta V^{\mathrm{PR}}_{\mathrm{oc}}$ and in turn $V_{\mathrm{oc}}$. Such small percentage of parasitic reabsorption probability lead to a $\sim 20$~mV reduction of $\Delta V^{\mathrm{PR}}_{\mathrm{oc}}$ at $Q_{\mathrm{i}}^{\mathrm{lum}}=1$ [ Fig. \ref{Fig5}b]. Note, that upon introduction of parasitic reabsorption, the increase of $\Delta V^{\mathrm{PR}}_{\mathrm{oc}}$ as $Q_{\mathrm{i}}^{\mathrm{lum}}$ approaches 1 also  considerably weaken. This is due to the parasitic reabsorption probability heavily impacting the possibility of multiple reemission and reabsorption events. The negative impact of parasitic absorption on photon recycling is more apparent when there is significant angular filtering as will be shown later.

\subsection{Angular dependence of absorption response}

\begin{figure}[h]
	\includegraphics[scale=0.23]{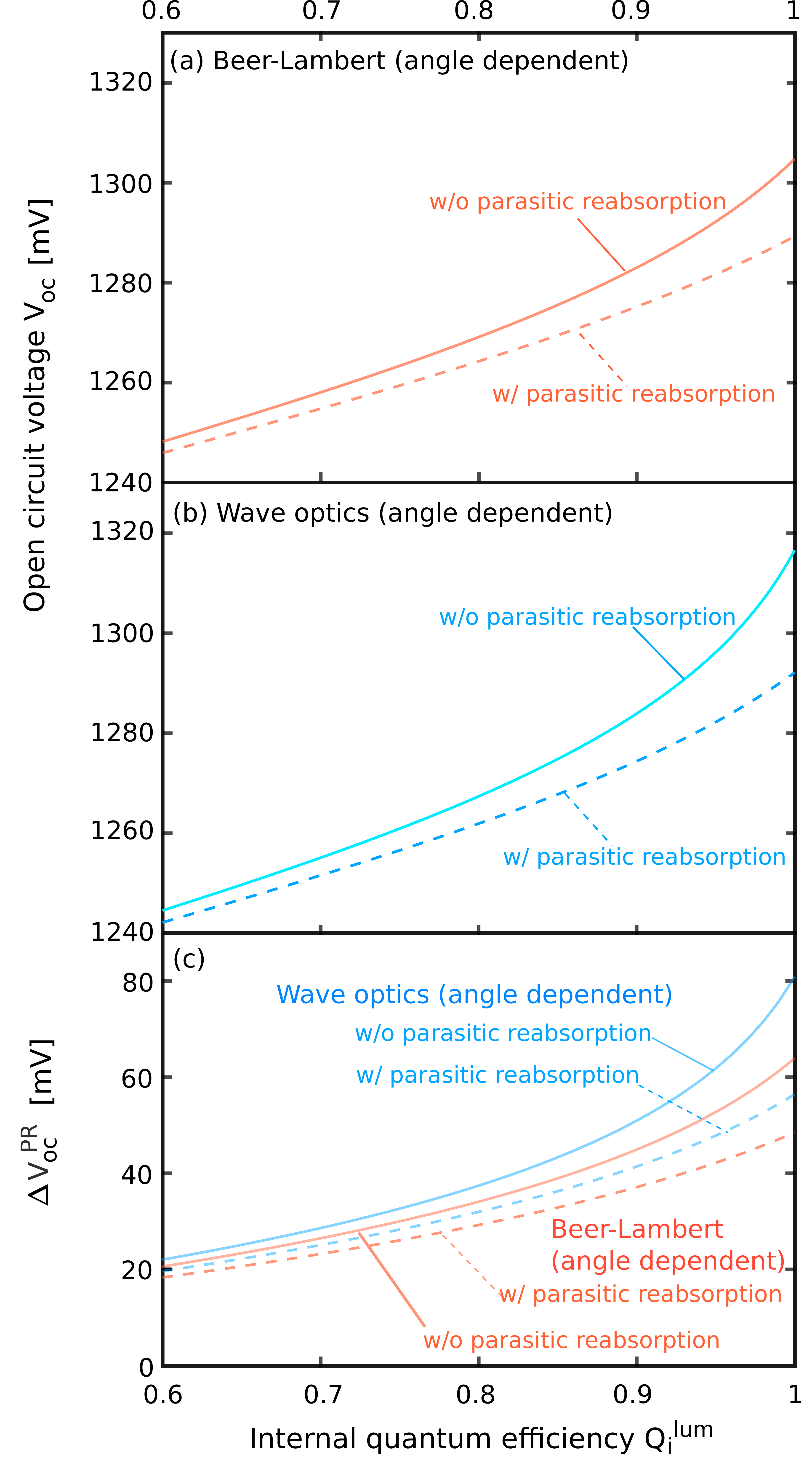}
	\caption{ $V_{\mathrm{oc}}$ as a function $Q^{\mathrm{lum}}_{\mathrm{i}}$ for (a) Beer-Lambert and (b) a wave optics angle dependent absorption responses. (c) The corresponding $\Delta V^{\mathrm{PR}}_{\mathrm{oc}}$ for both absorption cases. All plots are calculated assuming no angular restriction, with and without parasitic reabsorption being present due to other layers in the solar cell. \label{Fig6}}
\end{figure}

Having studied the impact of different light trapping schemes and realistic parasitic reabsorption probability, in this subsection we show the impact of considering an actual angle dependent absorption $A(E,\theta)$. We evaluate Eq.~\ref{eq_pebb} while considering the full wave-optical response of the reference stack and the Beer-Lambert case to deduce the escape probability $p_{\mathrm{e,bb}}$ with no angular restriction ($\theta_{\mathrm{out}}=\pi/2$). Accounting for angle dependent absorption response for the full planar stack and Beer-Lambert case leads to a notable reduction in  $V_{\mathrm{oc}}$ and $\Delta V^{\mathrm{PR}}_{\mathrm{oc}}$, as apparent from comparing the values in Figs.~\ref{Fig5} and \ref{Fig6}. A maximum open-circuit voltage value of $V_{\mathrm{oc}}$ of 1305 mV and 1317 mV is predicted for the Beer-Lambert and the full-stack case, respectively, when considering an angle dependent absorption in the absence of parasitic reabsorption.
The reduction of $V_{\mathrm{oc}}$ [Fig. \ref{Fig6}a] is due to a reduction of the photon recycling impact $\Delta V^{\mathrm{PR}}_{\mathrm{oc}}$ [Fig. \ref{Fig6}b]. A significant reduction in $\Delta V^{\mathrm{PR}}_{\mathrm{oc}}$ by $\sim 10-15$ mV relative to the angle independent response at $Q_{\mathrm{i}}^{\mathrm{lum}} = 1$ is apparent for both absorption cases. This is mainly due to the increase of absorption at large incidence angles, which translates to stronger emission in these angles as well.  One should note that a stronger discrepancy in $\Delta V^{\mathrm{PR}}_{\mathrm{oc}}$ values between angle dependent and independent absorption cases can occur if one considers a certain periodic light trapping or concentrating optics at the solar cell. There the absorption also greatly changes with angle of incidence. 

\subsection{Accurate escape probability obtained from rigorous dipole calculations}

\begin{figure}
	\includegraphics[scale=0.25]{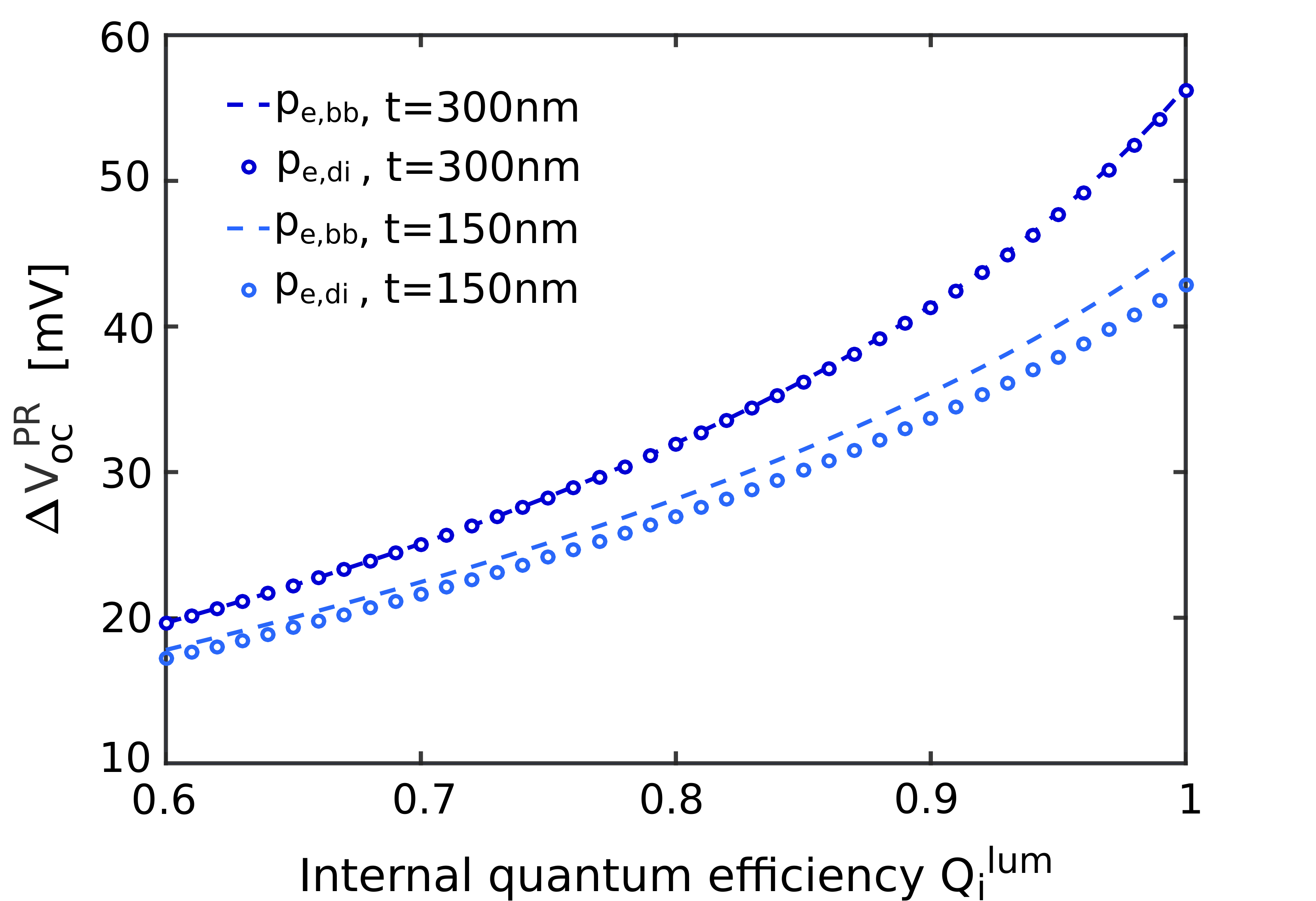}
	\caption{$\Delta V^{\mathrm{PR}}_{\mathrm{oc}}$ for the case of perovskite absorber thickness of $t=300$~nm and $t=150$~nm in the presence of parasitic absorption ($p_{\mathrm{a}}^{\mathrm{300nm}} = 0.0690$ and $p_{\mathrm{a}}^{\mathrm{150nm}}= 0.1282$).  We compare the obtained result when considering the escape probability calculated assuming a blackbody emission ($p_{\mathrm{e,bb}}^{\mathrm{300nm}} = 0.0437$ and $p_{\mathrm{e,bb}}^{\mathrm{150nm}}= 0.0426$) as compared to a rigorous analytical dipole calculation of the planar stack ($p_{\mathrm{e,di}}^{\mathrm{300nm}}=0.0448$ and $p_{\mathrm{e,di}}^{\mathrm{150nm}}=0.0623$). \label{Fig7}}
\end{figure}

In subsection C,  we considered the impact of light trapping, accurate parasitic reabsorption, and angle dependent absorption in calculating Eq.~\ref{eq_pebb} to deduce $p_{\mathrm{e,bb}}$. Here, we go a step further in accuracy by
utilizing Eq.~\ref{eq_averagepe} to deduce the reemitted photon escape probability ($p_{\mathrm{e,di}}$) for the planar multi-layer stack.
The rigorous dipole emission calculation (Eq.~\ref{eq_averagepe}) is found to be in good agreement with the black-body emission theory in predicting the escape probabilities (Eq. \ref{eq_pebb}). For perovskite absorber thickness of $t=300$ nm, we obtain $p_{\mathrm{e,bb}}=0.0437$ from black-body calculations considering angle dependent absorption (Eq.~\ref{eq_pebb}). The rigorous dipole calculation (Eq.~\ref{eq_averagepe}) gives $p_{\mathrm{e,di}}=0.0448$. As there is close agreement on the value of $p_{\mathrm{e}}$ for both approaches, also the deduced $\Delta V^{\mathrm{PR}}_{\mathrm{oc}}$ are in close agreement [Fig. \ref{Fig7}]. 
For other perovskite absorber thickness such as $t=150$ nm, a notable difference may occur where $p_{\mathrm{e,bb}}=0.0426$ and $p_{\mathrm{e,di}}=0.0623$. However, this only leads to a difference in $\Delta V^{\mathrm{PR}}_{\mathrm{oc}}$ up to a maximum of $\sim 2$ mV between both cases at $Q^{\mathrm{lum}}_{\mathrm{i}} = 1$ [Fig. \ref{Fig7}] due to the stronger parasitic reabsorption for $t=150$~nm. 

We wish to stress that only planar solar cells were considered herein. If one employs nanostructured solar cell architectures that either comprise of ordered or disordered scattering structures, the discrepancy of the escape probability values deduced with Eq.~\ref{eq_pebb} and Eq. \ref{eq_averagepe} can be even larger.

\subsection{Angular restriction}

Having considered the most accurate description of photon recycling for our considered solar cell device stack, we now examine photon recycling under "idealized" angular restriction where the escape cone $\theta_{\mathrm{out}}$ is changed without impacting other quantities such as the absorption response. Due to the nature of the angular restriction assumption, we evaluate the photon escape probability utilizing Eq.~\ref{eq_pebb} as the rigorous dipole treatment of Eq.~\ref{eq_averagepe} would require details of the angular restricting structure. 

\begin{figure}[h]
	\includegraphics[scale=0.255]{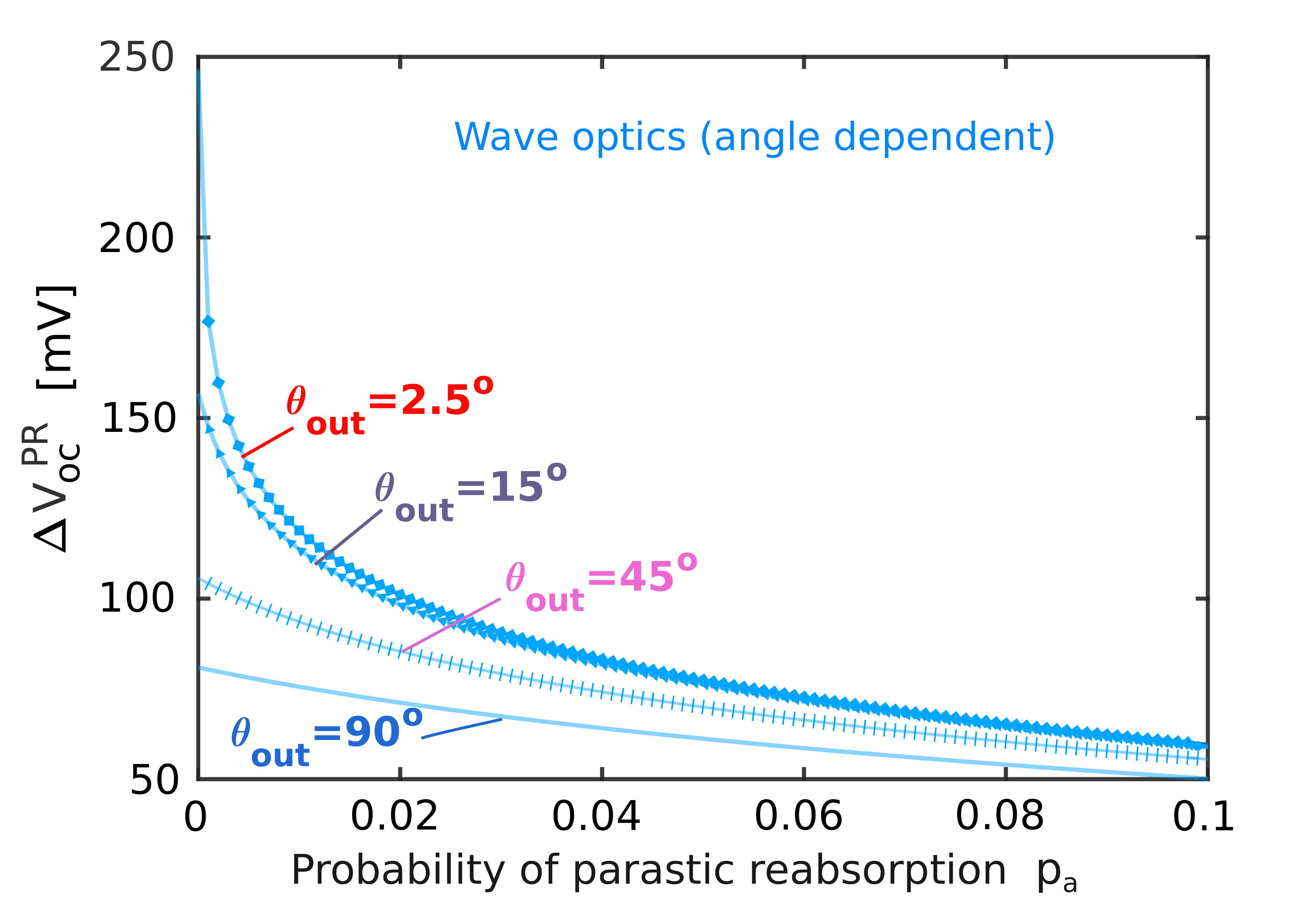}
	\caption{$\Delta V^{\mathrm{PR}}_{\mathrm{oc}}$ as a function of $p_{\mathrm{a}}$ at $Q^{\mathrm{lum}}_{\mathrm{i}} = 1$ for different angular filtering conditions ($\theta_{\mathrm{out}}=2.5^\circ-90^\circ$). \label{Fig8}}
\end{figure}

In order to strongly benefit from photon recycling through angular restriction, we find that one must maintain a very low $p_{\mathrm{a}}$. This is depicted in Fig. \ref{Fig8}, where we examine the dependency of $\Delta V^{\mathrm{PR}}_{\mathrm{oc}}$ on $p_{\mathrm{a}}$ at $Q^{\mathrm{lum}}_{\mathrm{i}} = 1$ and varying $\theta_{\mathrm{out}}$.
In the absence of parasitic reabsorption, one can potentially increase $\Delta V^{\mathrm{PR}}_{\mathrm{oc}}$ from 80 mV to 240 mV as one reduces the escape angle through angular filtering from $\theta_{\mathrm{out}}=90^\circ$ to $\theta_{\mathrm{out}}=2.5^\circ$. The increase in $\Delta V^{\mathrm{PR}}_{\mathrm{oc}}$ here is purely due to reemitted photons having less probability of escaping due to the smaller escape cone.  Fully restricting the escape cone to only $\theta_{\mathrm{out}}=\theta_{\mathrm{sun}}$ in the normal direction would lead to $\Delta V^{\mathrm{PR}}_{\mathrm{oc}} = 277$ mV for $p_{\mathrm{a}}=0$. This value corresponds to the etendue loss one expect in the absence of angular restriction as discussed by Rau \textit{et al.}~\cite{Rau2014}. This potential high $\Delta V^{\mathrm{PR}}_{\mathrm{oc}}$  at smaller $\theta_{\mathrm{out}}$ disappears with a small increase of $p_{\mathrm{a}}$. With just an increase of $p_{\mathrm{a}}$ from 0 to 0.02, one only gains $\sim25$~mV instead of  $\sim160$~mV by reducing $\theta_{\mathrm{out}}=90^\circ$ to $\theta_{\mathrm{out}}=2.5^\circ$.  Thus, to enhance photon recycling from angular restriction, one must maintain an extremely low parasitic photon reabsorption probability below 2$\%$ ($p_{\mathrm{a}} \leq 0.02$). Reaching such low values of $p_{\mathrm{a}}$ is indeed challenging as our rigorous calculation for the stack of Fig.~\ref{Fig2} already leads to $p_{\mathrm{a}} = 0.069$ for a commonly chosen perovskite layer thickness of $t=300$~nm. The thickness of the absorber layer greatly influences $p_{\mathrm{a}}$  as it determines the available optical modes in the system that reemitted photons can couple to. Additionally, thicker devices can naturally support a longer optical path within the absorber layer and in turn offer lower values of $p_{\mathrm{a}}$. In Fig.~\ref{AppC} of appendix C, we give the spatially resolved $p_{\mathrm{a}}$  for two planar perovskite solar cells with different thicknesses ($t$). $p_{\mathrm{a}}$ strongly oscillates along the device thickness due to wave-optical effects. For the solar cell architecture we consider, $p_{\mathrm{a}}$ is larger for thinner absorbers, purely due to the fact that there is less absorber material. The thinner device also exhibits a more strong spatial dependence due to resonant wave-optical effects. 

The harsh requirement on $p_{\mathrm{a}}$ is connected to the fact that the voltage enhancement one obtains through severe angular restriction relies on the possibility of having a high number of reemission and reabsorption events before carrier extraction or a photon escaping event. If recombination is dominantly radiative and reemission dominantly lead to reabsorption in the absorber, one naturally increases the open-circuit voltage and in turn the maximum power point voltage. With the introduction of a small parasitic reabsorption probability, one can greatly reduce the probability of having multiple reemission events prior to carrier collection and thereby severely limiting the impact of photon recycling.

\section{Summary}

In this work, we extend the theoretical framework on thermodynamics of photon management in solar cells to rigorously incorporate wave-optical effects. Our framework is valid for all single-junction solar cells electronically describable with semi-classical bulk semiconductor physics, encompassing thin-films and architectures with light trapping nanostructures, but applied here for the case of an organo-metal-halide thin-film perovskite cell. It allows to analyse open-circuit voltage ($V_{\mathrm{oc}}$) and the open-circuit voltage gain due to photon recycling $\Delta V^{\mathrm{PR}}_{\mathrm{oc}}$ of different accuracy levels, from rough estimations based on ray-optics to a fully rigorous treatment based on wave optics. A key result is provided by Eq.~\ref{eq_DVOCPR}, which shows the dependence of $\Delta V^{\mathrm{PR}}_{\mathrm{oc}}$ on internal quantum luminescence efficiency ($Q_{\mathrm{i}}^{\mathrm{lum}}$) and probabilities of parasitic reabsorption probability ($p_\mathrm{a}$) and escape probability ($p_\mathrm{e}$) of reemitted photons. Exploiting our analytical expressions, we depict different regimes of $Q^{\mathrm{lum}}_{\mathrm{i}} $ with varying impact of photon recycling on $V_{\mathrm{oc}}$. A $Q^{\mathrm{lum}}_{\mathrm{i}} $ of near unity is paramount along with an extremely small probability for parasitic reabsorption. 

For the exemplary case of a conventional perovskite thin-film solar cells, in the absence of angular restriction and parasitic reabsorption, a voltage enhancement  $\Delta V^{\mathrm{PR}}_{\mathrm{oc}}$ of 80 mV is determined with our fully rigorous analysis, confirming the $\Delta V^{\mathrm{PR}}_{\mathrm{oc}}$ determined in previous reports based on non rigorous analyses. There can be discrepancy of the photon escape probability ($p_\mathrm{e}$) calculated by the non-rigorous ray-optics based approach (Eq.~\ref{eq_pebb}) and our rigorous full wave optical calculation (Eq.~\ref{eq_averagepe}) depending on the thickness of the considered cell, which in turn lead to a significant difference in the predicted $V_{\mathrm{oc}}$ and $\Delta V^{\mathrm{PR}}_{\mathrm{oc}}$ should parasitic reabsorption be small enough. The need of rigorous treatment of photon recycling would be more paramount in solar cell architectures which are heavily impacted by wave optics, such as nanowires or other nanostructured solar cells. We also show that under angular restriction of $\theta_{\mathrm{out}}=2.5^\circ$, one could potentially obtain $\Delta V^{\mathrm{PR}}_{\mathrm{oc}}$ of 240 mV with the considered perovskite layer stack. The high $\Delta V^{\mathrm{PR}}_{\mathrm{oc}}$ reduces by more than half to $\sim$100~ mV by just an increase of parasitic reabsorption probability to only 2$\%$.  

The rigorous optical treatment of photon recycling in a nano-patterned solar cells (exploiting e.g. front side diffraction grating) is the object of a future study, which will account for the exact and spatially-averaged emission and parasitic reabsorption probabilities.

\section{Acknowledgement}
The authors acknowledge support by the Helmholtz Association – through the program “Science and Technology of Nanosystems (STN)”, the Karlsruhe Nano Micro Facility (KNMF), Helmholtz Postdoctoral Program (G. Gomard), the Funding of the Helmholtz Association (HYIG of U. Paetzold and Recruitment Initiative of B. S. Richards), Initiating and Networking Funding of the Helmholtz Association (PEROSEED), the KIT Young Investigator Network, and the Karlsruhe School of Optics and Photonics (KSOP). This project has received funding from the EMPIR programme co-financed
by the Participating States and from the European Union’s Horizon 2020 research and innovation programme under grant agreement number 14IND13 (PhotInd).

\section{Appendix}

\subsection{Open-circuit voltage derivation details}

We begin with the DB open-circuit voltage expression in its extended form accounting for non-radiative recombination \cite{Roosbroeck1954,Rau2014}  
	\begin{equation}\label{xi}
qV_{\mathrm{oc}}=qV_{\mathrm{oc}}^{\mathrm{rad}}+kT_{c}{\ln\left\lbrace Q_{\mathrm{e}}^{\mathrm{LED}}\right\rbrace}.
	\end{equation}
$Q_{\mathrm{e}}^{\mathrm{LED}}$ is described by
\begin{equation}\label{EQE}
Q_{\mathrm{e}}^{\mathrm{LED}}=p_{\mathrm{e}}R^{\mathrm{rad}}_{\mathrm{int}}/(R^{\mathrm{total}}).
\end{equation}
where $R^{\mathrm{total}}=(1-p_{\mathrm{r}})R^{\mathrm{rad}}_{\mathrm{int}}+R^{\mathrm{nrd}}$ is the total recombination rate with $p_{\mathrm{r}}$ is the probability of reemitted photons being reabsorbed in the absorber layer, thus $p_{\mathrm{r}}+p_{\mathrm{a}}+p_{\mathrm{e}}=1$ .
$R^{\mathrm{nrd}}$ is the non-radiative recombination rate, which results in the non-radiative saturation current responsible for loss ( $J_{\mathrm{re,nrd}}=qR^{\mathrm{nrd}}$). $R^{\mathrm{rad}}_{\mathrm{int}}$ is the internal radiative recombination rate.

$Q_{\mathrm{i}}^{\mathrm{lum}}$ is a pure internal property which describes how much portion of the recombination is radiative. We make the simplifying assumption here that the recombination rates, and therefore , at operating and open circuit bias voltage to be approximately equivalent. This is applicable for low injection regimes in semiclassical semiconductor bulks. $Q_{\mathrm{i}}^{\mathrm{lum}}$ is therefore describable by  $Q_{\mathrm{i}}^{\mathrm{lum}}=R^{\mathrm{rad}}_{\mathrm{int}}/(R^{\mathrm{rad}}_{\mathrm{int}}+R^{\mathrm{nrd}})$.
One can therefore reformulate the non-radiative recombination rate as
\begin{equation}\label{Rnrd}
R^{\mathrm{nrd}}=R^{\mathrm{rad}}_{\mathrm{int}}\left[\frac{1}{Q_{\mathrm{i}}^{\mathrm{lum}}}-1\right].
\end{equation}
Utilizing $p_{\mathrm{e}}+p_{\mathrm{a}}+p_{\mathrm{r}}=1$, one can therefore write $Q_{\mathrm{e}}^{\mathrm{LED}}$ as
\begin{equation}\label{EQE3}
	Q_{\mathrm{e}}^{\mathrm{LED}}=\frac{p_{\mathrm{e}}}{ p_{\mathrm{e}}+p_{\mathrm{a}}+\left(\frac{1}{Q_{\mathrm{i}}^{\mathrm{lum}}}-1\right)}.
	\end{equation}

Utilizing  $p_{\mathrm{e}}=J_{\mathrm{em}}/ J_{\mathrm{re,rad}}$ together with Eqs.~\ref{eq_VocSQ} and \ref{EQE3}, one can reformulate Eq.~\ref{xi} to write the open-circuit voltage in the form of Eq. \ref{eq_VocDB} as shown in the main text.	

We wish to further clarify a confusion in the community concerning the belief that $V_{\mathrm{oc}}$ can be enhanced by increasing outcoupling efficiency, which essentially implies increasing  $p_{\mathrm{e}}$ and thereby $Q_{\mathrm{e}}^{\mathrm{LED}}$. Enhancement of $Q_{\mathrm{e}}^{\mathrm{LED}}$, however, does not necessarily lead to enhancement of $V_{\mathrm{oc}}$. To show this clearly, we note that $V_{\mathrm{oc}}^{\mathrm{rad}}$ can also be expressed in the form
\begin{equation}\label{VocR}
	qV_{\mathrm{oc}}^{\mathrm{rad}}=kT_{c}{\ln\left\lbrace \frac{J_{\mathrm{sun}}}{p_{\mathrm{e}}J_{\mathrm{re,rad}}}\right\rbrace}.
\end{equation}
By comparing Eqs.~\ref{VocR} and \ref{EQE3}, one can see that the $V_{\mathrm{oc}}^{\mathrm{rad}}$ is not independent to $Q_{\mathrm{e}}^{\mathrm{LED}}$. In particular, the photon escape probability $p_{\mathrm{e}}$, which depends on the absorption of the solar cell, impacts both. One cannot therefore expect that increasing outcoupling efficiency always lead to $V_{\mathrm{oc}}$ enhancement. In fact, an increase in $p_{\mathrm{e}}$ without a compensating increase in $J_{\mathrm{sun}}$ can greatly reduce $V_{\mathrm{oc}}$ as well. 

\subsection{Absorption response}

\begin{figure}[H]
	\includegraphics[scale=0.25]{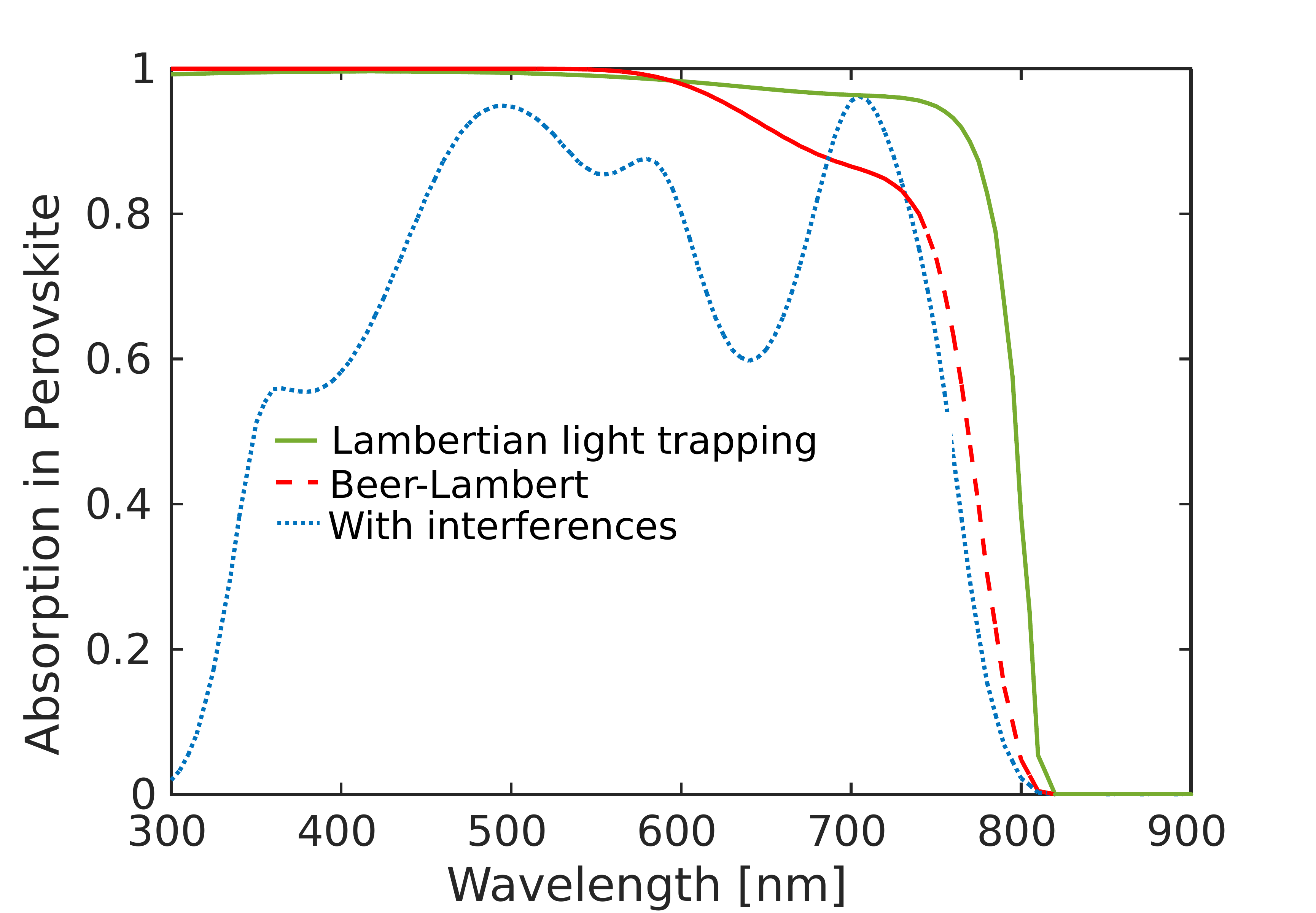}
	\caption{Absorption spectra of an organo-metal halide Perovskite cell with a Lambertian front texture, a perfect anti-reflection only, and for a full device stack as depicted in Fig. \ref{Fig2}. All absorption calculations are for the perovskite layer thickness of 300~nm. \label{AppB}}
\end{figure}

We give in Fig. \ref{AppB} a plot of the three considered absorption response at normal incidence for the case of a 300~nm thick organo-metal halide perovskite absorber layer. 

\subsection{The dipole emission calculation details}

The rigorous plane wave expansion of the Green's tensor was done for the case of a semi-infinite planar multilayer organo-metal halide Perovskite solar cell device stack of Fig. \ref{Fig2} for the wavelength of 770~nm, which is the peak emission wavelength for the organo-metal halide perovskite layer we consider\cite{Staub2016}. We calculated for both dipole orientations parallel and perpendicular to the layer stack and for different positions along the thickness of the perovskite layer. 

We wish to note that classical dipole emission calculations in a dissipative media, as what is needed here, requires the emitter to be enclosed in a small non-absorbing cavity to ensure a well defined radiated power \cite{Tai2000}. This is fundamentally tied to the introduction of additional non-radiative decay channels when the emitter is embedded or is extremely close to absorbing media, which thus require extra care to properly treat it \cite{Barnett1992,Schell1999,Juzeliunas2006}. Calculations utilizing a small non-absorbing cavity has been shown to correspond to experimental observation \cite{Rikken1995,Schuurmans1998}. Considering dipole emission directly within the perovskite absorber leads to numerical artifacts where energy is not conserved. To avoid this problem, we introduce a thin non-absorbing layer with the same real part refractive index as the perovskite layer which surrounds the dipole emitter. The introduction of this non-absorbing layer results in a slight overestimation of $p_\mathrm{a}$ as reabsorption within the perovskite layer will be underestimated.

In calculating the parasitically reabsorbed portion of the power, we calculate the difference of power flux through the top and bottom boundaries of each layer and then performing a sum for all considered layers and normalizing the result to  $P^{\mathrm{dip}}_{o}$. Calculating the power flux at the interfaces of each layer, as we described, is essentially equivalent to calculating the integral in Eq.~\ref{eq_averagepa} and allow ease in ensuring numerical accuracy. $A^\mathrm{parasitic}_o$ can be mathematically expressed by 
\begin{equation}
A^\mathrm{parasitic}_o=\frac{\sum_l P^\mathrm{top}_{o,l}-P^\mathrm{bottom}_{o,l} } {P^{\mathrm{dip}}_{o}},
\end{equation}.
where the index $l$ labels the different layers considered, $P_{\mathrm{top},l}$ and $P_{\mathrm{bottom},l}$ are the power flux at the top and bottom interface of each layer, respectively.

Calculating the parasitic absorption in this manner would require us to consider a large computational domain in order to accurately calculate $P_{\mathrm{top},l}$ and $P_{\mathrm{bottom},l}$, especially when modes with long propagation lengths are involved. It is therefore important to also calculate the portion of the emitted power absorbed in the perovskite($A^\mathrm{perovskite}$) and the portion of emitted power that leaves the layer stack ($\tilde{P}^\mathrm{esc}$). In our calculations, we ensured the computational domain to be wide enough such that all power are accounted for within the computational domain $A^\mathrm{parasitic}+A^\mathrm{perovskite}+\tilde{P}^\mathrm{esc}=1$. 

The non-absorbing layer surrounding the dipole emitter is chosen in our calculations to be sufficiently thick to avoid the interaction of the dipole near-field components with the surrounding absorbing media, which leads to an incorrect quantification of the dipole emitted power with our classical approach. From numerical experiments, we found that energy conservation is maintained when we utilized a non-absorbing layer with thickness $\geq 40$nm in which the dipole emitter is placed at the center. We thus employ a non-absorbing layer with a total thickness of 40~nm for all our dipole emission calculations. Increasing or reducing the non-absorbing layer thickness by $\pm 4$~nm ($\pm10\%$) leads to a change in parasitic absorption by $\Delta p_\mathrm{a} = \mp0.005$ for perovskite thickness $t=300$~nm and $\Delta p_\mathrm{a} = \mp 0.01$ for  $t=150$~nm. The escape probability, on the other hand changes by $\Delta p_\mathrm{e} = \mp0.003$ for perovskite thickness $t=300$~nm and $\Delta p_\mathrm{a} = \mp 0.004$ for $t=150$~nm. As the change of the values of $p_\mathrm{a}$ and $p_\mathrm{e}$ is not overly sensitive to the change of the non-absorbing layer thickness, we thus believe our dipole calculations provide a sufficiently accurate calculations of the values of $p_\mathrm{a}$ and $p_\mathrm{e}$ for the perovskite absorber thicknesses we consider.

\begin{figure}
	\includegraphics[scale=0.25]{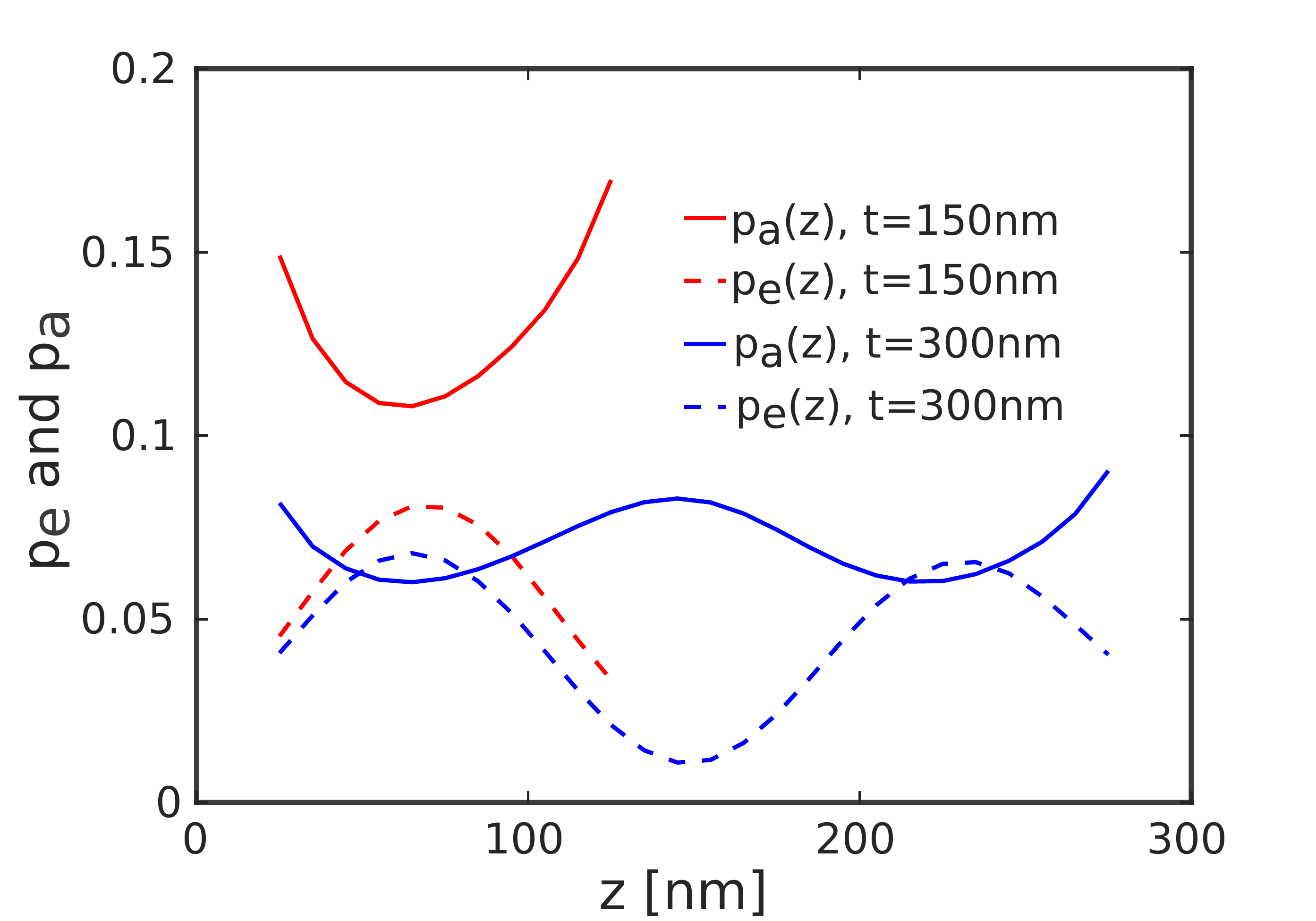}
	\caption{Spatial dependence of the parasitic photon reabsorption and the escape photon probabilities along the thickness. Both probability values are weighted average quantities for all possible dipole emission orientations as calculated by Eq.~\ref{eq_weightedAv}. \label{AppC}}
\end{figure}

After obtaining $A_{\mathrm{parasitic},o}$ for every dipole orientation at a certain position, we then calculate the averaged parasitic reabsorption probability at a particular height $z$ along the thickness $p_\mathrm{a}(z_0)$ with 
\begin{equation}
p_\mathrm{a}(z)=\frac{\sum_o P_{\mathrm{dip},o}(z_0) A_{\mathrm{parasitic},o} } {\sum_o P_{\mathrm{dip},o}(z)}\label{eq_weightedAv},
\end{equation}
where the index $o$ indicates the cartesian directions \textit{xyz}. Equation~\ref{eq_weightedAv} is essentially equivalent to the righthand side of Eq.~\ref{eq_averagepa}, without the spatial average. A plot of the spatially resolved ${p}_\mathrm{a}(z)$ and ${p}_\mathrm{a}(e)$ for the case of the layer stack in Fig.~\ref{Fig2} is given in Fig. \ref{AppC} assuming $t=300$~nm and $t=150$~nm.

\bibliographystyle{apsrev4-1}
\bibliography{References}

\begin{thebibliography}{52}%
\makeatletter
\providecommand \@ifxundefined [1]{%
 \@ifx{#1\undefined}
}%
\providecommand \@ifnum [1]{%
 \ifnum #1\expandafter \@firstoftwo
 \else \expandafter \@secondoftwo
 \fi
}%
\providecommand \@ifx [1]{%
 \ifx #1\expandafter \@firstoftwo
 \else \expandafter \@secondoftwo
 \fi
}%
\providecommand \natexlab [1]{#1}%
\providecommand \enquote  [1]{``#1''}%
\providecommand \bibnamefont  [1]{#1}%
\providecommand \bibfnamefont [1]{#1}%
\providecommand \citenamefont [1]{#1}%
\providecommand \href@noop [0]{\@secondoftwo}%
\providecommand \href [0]{\begingroup \@sanitize@url \@href}%
\providecommand \@href[1]{\@@startlink{#1}\@@href}%
\providecommand \@@href[1]{\endgroup#1\@@endlink}%
\providecommand \@sanitize@url [0]{\catcode `\\12\catcode `\$12\catcode
  `\&12\catcode `\#12\catcode `\^12\catcode `\_12\catcode `\%12\relax}%
\providecommand \@@startlink[1]{}%
\providecommand \@@endlink[0]{}%
\providecommand \url  [0]{\begingroup\@sanitize@url \@url }%
\providecommand \@url [1]{\endgroup\@href {#1}{\urlprefix }}%
\providecommand \urlprefix  [0]{URL }%
\providecommand \Eprint [0]{\href }%
\providecommand \doibase [0]{http://dx.doi.org/}%
\providecommand \selectlanguage [0]{\@gobble}%
\providecommand \bibinfo  [0]{\@secondoftwo}%
\providecommand \bibfield  [0]{\@secondoftwo}%
\providecommand \translation [1]{[#1]}%
\providecommand \BibitemOpen [0]{}%
\providecommand \bibitemStop [0]{}%
\providecommand \bibitemNoStop [0]{.\EOS\space}%
\providecommand \EOS [0]{\spacefactor3000\relax}%
\providecommand \BibitemShut  [1]{\csname bibitem#1\endcsname}%
\let\auto@bib@innerbib\@empty
\bibitem [{\citenamefont {Parrott}(1993)}]{Parrott1993}%
  \BibitemOpen
  \bibfield  {author} {\bibinfo {author} {\bibfnamefont {J.}~\bibnamefont
  {Parrott}},\ }\href {\doibase https://doi.org/10.1016/0927-0248(93)90142-P}
  {\bibfield  {journal} {\bibinfo  {journal} {Sol. Energy Mater. Sol. Cells}\
  }\textbf {\bibinfo {volume} {30}},\ \bibinfo {pages} {221 } (\bibinfo {year}
  {1993})}\BibitemShut {NoStop}%
\bibitem [{\citenamefont {Badescu}\ and\ \citenamefont
  {Landsberg}(1997)}]{Badescu1997}%
  \BibitemOpen
  \bibfield  {author} {\bibinfo {author} {\bibfnamefont {V.}~\bibnamefont
  {Badescu}}\ and\ \bibinfo {author} {\bibfnamefont {P.~T.}\ \bibnamefont
  {Landsberg}},\ }\href {http://stacks.iop.org/0268-1242/12/i=11/a=028}
  {\bibfield  {journal} {\bibinfo  {journal} {Semicond. Sci. Technol.}\
  }\textbf {\bibinfo {volume} {12}},\ \bibinfo {pages} {1491} (\bibinfo {year}
  {1997})}\BibitemShut {NoStop}%
\bibitem [{\citenamefont {Zeng}\ \emph {et~al.}(2016)\citenamefont {Zeng},
  \citenamefont {Xue},\ and\ \citenamefont {Tang}}]{Zeng2016}%
  \BibitemOpen
  \bibfield  {author} {\bibinfo {author} {\bibfnamefont {K.}~\bibnamefont
  {Zeng}}, \bibinfo {author} {\bibfnamefont {D.-J.}\ \bibnamefont {Xue}}, \
  and\ \bibinfo {author} {\bibfnamefont {J.}~\bibnamefont {Tang}},\ }\href
  {http://stacks.iop.org/0268-1242/31/i=6/a=063001} {\bibfield  {journal}
  {\bibinfo  {journal} {Semicond. Sci. Technol.}\ }\textbf {\bibinfo {volume}
  {31}},\ \bibinfo {pages} {063001} (\bibinfo {year} {2016})}\BibitemShut
  {NoStop}%
\bibitem [{\citenamefont {Kirchartz}\ \emph {et~al.}(2016)\citenamefont
  {Kirchartz}, \citenamefont {Staub},\ and\ \citenamefont
  {Rau}}]{Kirchatz2016}%
  \BibitemOpen
  \bibfield  {author} {\bibinfo {author} {\bibfnamefont {T.}~\bibnamefont
  {Kirchartz}}, \bibinfo {author} {\bibfnamefont {F.}~\bibnamefont {Staub}}, \
  and\ \bibinfo {author} {\bibfnamefont {U.}~\bibnamefont {Rau}},\ }\href
  {\doibase 10.1021/acsenergylett.6b00223} {\bibfield  {journal} {\bibinfo
  {journal} {ACS Energy Lett.}\ }\textbf {\bibinfo {volume} {1}},\ \bibinfo
  {pages} {731} (\bibinfo {year} {2016})}\BibitemShut {NoStop}%
\bibitem [{\citenamefont {van Roosbroeck}\ and\ \citenamefont
  {Shockley}(1954)}]{Roosbroeck1954}%
  \BibitemOpen
  \bibfield  {author} {\bibinfo {author} {\bibfnamefont {W.}~\bibnamefont {van
  Roosbroeck}}\ and\ \bibinfo {author} {\bibfnamefont {W.}~\bibnamefont
  {Shockley}},\ }\href {\doibase 10.1103/PhysRev.94.1558} {\bibfield  {journal}
  {\bibinfo  {journal} {Phys. Rev.}\ }\textbf {\bibinfo {volume} {94}},\
  \bibinfo {pages} {1558} (\bibinfo {year} {1954})}\BibitemShut {NoStop}%
\bibitem [{\citenamefont {Dumke}(1957)}]{Dumke1957}%
  \BibitemOpen
  \bibfield  {author} {\bibinfo {author} {\bibfnamefont {W.~P.}\ \bibnamefont
  {Dumke}},\ }\href {\doibase 10.1103/PhysRev.105.139} {\bibfield  {journal}
  {\bibinfo  {journal} {Phys. Rev.}\ }\textbf {\bibinfo {volume} {105}},\
  \bibinfo {pages} {139} (\bibinfo {year} {1957})}\BibitemShut {NoStop}%
\bibitem [{\citenamefont {Smestad}\ and\ \citenamefont
  {Ries}(1992)}]{Smestad1992}%
  \BibitemOpen
  \bibfield  {author} {\bibinfo {author} {\bibfnamefont {G.}~\bibnamefont
  {Smestad}}\ and\ \bibinfo {author} {\bibfnamefont {H.}~\bibnamefont {Ries}},\
  }\href {\doibase 10.1016/0927-0248(92)90016-I} {\bibfield  {journal}
  {\bibinfo  {journal} {Solar Energy Materials and Solar Cells}\ }\textbf
  {\bibinfo {volume} {25}},\ \bibinfo {pages} {51} (\bibinfo {year}
  {1992})}\BibitemShut {NoStop}%
\bibitem [{\citenamefont {Green}(2012)}]{Green2012}%
  \BibitemOpen
  \bibfield  {author} {\bibinfo {author} {\bibfnamefont {M.~A.}\ \bibnamefont
  {Green}},\ }\href {\doibase 10.1002/pip.1147} {\bibfield  {journal} {\bibinfo
   {journal} {Prog. Photovolt. Res. Appl.}\ }\textbf {\bibinfo {volume} {20}},\
  \bibinfo {pages} {472} (\bibinfo {year} {2012})}\BibitemShut {NoStop}%
\bibitem [{\citenamefont {Miller}\ \emph {et~al.}(2012)\citenamefont {Miller},
  \citenamefont {Yablonovitch},\ and\ \citenamefont {Kurtz}}]{Miller2012}%
  \BibitemOpen
  \bibfield  {author} {\bibinfo {author} {\bibfnamefont {O.~D.}\ \bibnamefont
  {Miller}}, \bibinfo {author} {\bibfnamefont {E.}~\bibnamefont
  {Yablonovitch}}, \ and\ \bibinfo {author} {\bibfnamefont {S.~R.}\
  \bibnamefont {Kurtz}},\ }\href {\doibase 10.1109/JPHOTOV.2012.2198434}
  {\bibfield  {journal} {\bibinfo  {journal} {IEEE J. Photovolt.}\ }\textbf
  {\bibinfo {volume} {2}},\ \bibinfo {pages} {303} (\bibinfo {year}
  {2012})}\BibitemShut {NoStop}%
\bibitem [{\citenamefont {Rau}\ \emph {et~al.}(2014)\citenamefont {Rau},
  \citenamefont {Paetzold},\ and\ \citenamefont {Kirchartz}}]{Rau2014}%
  \BibitemOpen
  \bibfield  {author} {\bibinfo {author} {\bibfnamefont {U.}~\bibnamefont
  {Rau}}, \bibinfo {author} {\bibfnamefont {U.~W.}\ \bibnamefont {Paetzold}}, \
  and\ \bibinfo {author} {\bibfnamefont {T.}~\bibnamefont {Kirchartz}},\ }\href
  {\doibase 10.1103/PhysRevB.90.035211} {\bibfield  {journal} {\bibinfo
  {journal} {Phys. Rev. B}\ }\textbf {\bibinfo {volume} {90}},\ \bibinfo
  {pages} {035211} (\bibinfo {year} {2014})}\BibitemShut {NoStop}%
\bibitem [{\citenamefont {Staub}\ \emph {et~al.}(2017)\citenamefont {Staub},
  \citenamefont {Kirchartz}, \citenamefont {Bittkau},\ and\ \citenamefont
  {Rau}}]{Staub2017}%
  \BibitemOpen
  \bibfield  {author} {\bibinfo {author} {\bibfnamefont {F.}~\bibnamefont
  {Staub}}, \bibinfo {author} {\bibfnamefont {T.}~\bibnamefont {Kirchartz}},
  \bibinfo {author} {\bibfnamefont {K.}~\bibnamefont {Bittkau}}, \ and\
  \bibinfo {author} {\bibfnamefont {U.}~\bibnamefont {Rau}},\ }\href {\doibase
  10.1021/acs.jpclett.7b02224} {\bibfield  {journal} {\bibinfo  {journal} {J.
  Phys. Chem. Lett.}\ }\textbf {\bibinfo {volume} {8}},\ \bibinfo {pages}
  {5084} (\bibinfo {year} {2017})}\BibitemShut {NoStop}%
\bibitem [{\citenamefont {Steiner}\ \emph {et~al.}(2013)\citenamefont
  {Steiner}, \citenamefont {Geisz}, \citenamefont {García}, \citenamefont
  {Friedman}, \citenamefont {Duda},\ and\ \citenamefont {Kurtz}}]{Steiner2013}%
  \BibitemOpen
  \bibfield  {author} {\bibinfo {author} {\bibfnamefont {M.~A.}\ \bibnamefont
  {Steiner}}, \bibinfo {author} {\bibfnamefont {J.~F.}\ \bibnamefont {Geisz}},
  \bibinfo {author} {\bibfnamefont {I.}~\bibnamefont {García}}, \bibinfo
  {author} {\bibfnamefont {D.~J.}\ \bibnamefont {Friedman}}, \bibinfo {author}
  {\bibfnamefont {A.}~\bibnamefont {Duda}}, \ and\ \bibinfo {author}
  {\bibfnamefont {S.~R.}\ \bibnamefont {Kurtz}},\ }\href@noop {} {\bibfield
  {journal} {\bibinfo  {journal} {J. Appl. Phys.}\ }\textbf {\bibinfo {volume}
  {113}},\ \bibinfo {pages} {123109} (\bibinfo {year} {2013})}\BibitemShut
  {NoStop}%
\bibitem [{\citenamefont {Braun}\ \emph {et~al.}(2013)\citenamefont {Braun},
  \citenamefont {Katz}, \citenamefont {Feuermann}, \citenamefont {Kayes},\ and\
  \citenamefont {Gordon}}]{Braun2013}%
  \BibitemOpen
  \bibfield  {author} {\bibinfo {author} {\bibfnamefont {A.}~\bibnamefont
  {Braun}}, \bibinfo {author} {\bibfnamefont {E.~A.}\ \bibnamefont {Katz}},
  \bibinfo {author} {\bibfnamefont {D.}~\bibnamefont {Feuermann}}, \bibinfo
  {author} {\bibfnamefont {B.~M.}\ \bibnamefont {Kayes}}, \ and\ \bibinfo
  {author} {\bibfnamefont {J.~M.}\ \bibnamefont {Gordon}},\ }\href {\doibase
  10.1039/C3EE40377G} {\bibfield  {journal} {\bibinfo  {journal} {Energy
  Environ. Sci.}\ }\textbf {\bibinfo {volume} {6}},\ \bibinfo {pages} {1499}
  (\bibinfo {year} {2013})}\BibitemShut {NoStop}%
\bibitem [{\citenamefont {Kosten}\ \emph {et~al.}(2014)\citenamefont {Kosten},
  \citenamefont {Kayes},\ and\ \citenamefont {Atwater}}]{Kosten2014}%
  \BibitemOpen
  \bibfield  {author} {\bibinfo {author} {\bibfnamefont {E.~D.}\ \bibnamefont
  {Kosten}}, \bibinfo {author} {\bibfnamefont {B.~M.}\ \bibnamefont {Kayes}}, \
  and\ \bibinfo {author} {\bibfnamefont {H.~A.}\ \bibnamefont {Atwater}},\
  }\href {\doibase 10.1039/C3EE43584A} {\bibfield  {journal} {\bibinfo
  {journal} {Energy Environ. Sci.}\ }\textbf {\bibinfo {volume} {7}},\ \bibinfo
  {pages} {1907} (\bibinfo {year} {2014})}\BibitemShut {NoStop}%
\bibitem [{\citenamefont {Vossier}\ \emph {et~al.}(2015)\citenamefont
  {Vossier}, \citenamefont {Gualdi}, \citenamefont {Dollet}, \citenamefont
  {Ares},\ and\ \citenamefont {Aimez}}]{Vossier2015}%
  \BibitemOpen
  \bibfield  {author} {\bibinfo {author} {\bibfnamefont {A.}~\bibnamefont
  {Vossier}}, \bibinfo {author} {\bibfnamefont {F.}~\bibnamefont {Gualdi}},
  \bibinfo {author} {\bibfnamefont {A.}~\bibnamefont {Dollet}}, \bibinfo
  {author} {\bibfnamefont {R.}~\bibnamefont {Ares}}, \ and\ \bibinfo {author}
  {\bibfnamefont {V.}~\bibnamefont {Aimez}},\ }\href {\doibase
  10.1063/1.4905277} {\bibfield  {journal} {\bibinfo  {journal} {J. Appl.
  Phys.}\ }\textbf {\bibinfo {volume} {117}},\ \bibinfo {pages} {015102}
  (\bibinfo {year} {2015})}\BibitemShut {NoStop}%
\bibitem [{\citenamefont {Walker}\ \emph {et~al.}(2015)\citenamefont {Walker},
  \citenamefont {Höhn}, \citenamefont {Micha}, \citenamefont {Bläsi},
  \citenamefont {Bett},\ and\ \citenamefont {Dimroth}}]{Walker2015}%
  \BibitemOpen
  \bibfield  {author} {\bibinfo {author} {\bibfnamefont {A.~W.}\ \bibnamefont
  {Walker}}, \bibinfo {author} {\bibfnamefont {O.}~\bibnamefont {Höhn}},
  \bibinfo {author} {\bibfnamefont {D.~N.}\ \bibnamefont {Micha}}, \bibinfo
  {author} {\bibfnamefont {B.}~\bibnamefont {Bläsi}}, \bibinfo {author}
  {\bibfnamefont {A.~W.}\ \bibnamefont {Bett}}, \ and\ \bibinfo {author}
  {\bibfnamefont {F.}~\bibnamefont {Dimroth}},\ }\href@noop {} {\bibfield
  {journal} {\bibinfo  {journal} {IEEE J. Photovolt.}\ }\textbf {\bibinfo
  {volume} {5}},\ \bibinfo {pages} {1636} (\bibinfo {year} {2015})}\BibitemShut
  {NoStop}%
\bibitem [{\citenamefont {Saliba}\ \emph {et~al.}(2015)\citenamefont {Saliba},
  \citenamefont {Zhang}, \citenamefont {Burlakov}, \citenamefont {Stranks},
  \citenamefont {Sun}, \citenamefont {Ball}, \citenamefont {Johnston},
  \citenamefont {Goriely}, \citenamefont {Wiesner},\ and\ \citenamefont
  {Snaith}}]{Saliba2015}%
  \BibitemOpen
  \bibfield  {author} {\bibinfo {author} {\bibfnamefont {M.}~\bibnamefont
  {Saliba}}, \bibinfo {author} {\bibfnamefont {W.}~\bibnamefont {Zhang}},
  \bibinfo {author} {\bibfnamefont {V.~M.}\ \bibnamefont {Burlakov}}, \bibinfo
  {author} {\bibfnamefont {S.~D.}\ \bibnamefont {Stranks}}, \bibinfo {author}
  {\bibfnamefont {Y.}~\bibnamefont {Sun}}, \bibinfo {author} {\bibfnamefont
  {J.~M.}\ \bibnamefont {Ball}}, \bibinfo {author} {\bibfnamefont {M.~B.}\
  \bibnamefont {Johnston}}, \bibinfo {author} {\bibfnamefont {A.}~\bibnamefont
  {Goriely}}, \bibinfo {author} {\bibfnamefont {U.}~\bibnamefont {Wiesner}}, \
  and\ \bibinfo {author} {\bibfnamefont {H.~J.}\ \bibnamefont {Snaith}},\
  }\href {\doibase 10.1002/adfm.201500669} {\bibfield  {journal} {\bibinfo
  {journal} {Adv. Funct. Mater.}\ }\textbf {\bibinfo {volume} {25}},\ \bibinfo
  {pages} {5038} (\bibinfo {year} {2015})}\BibitemShut {NoStop}%
\bibitem [{\citenamefont {NREL}(2018)}]{NREL}%
  \BibitemOpen
  \bibfield  {author} {\bibinfo {author} {\bibnamefont {NREL}},\ }\href
  {https://www.nrel.gov/pv/assets/images/efficiency-chart.png} {\enquote
  {\bibinfo {title} {Best research-cell efficiencies},}\ } (\bibinfo {year}
  {2018})\BibitemShut {NoStop}%
\bibitem [{\citenamefont {Stranks}\ \emph {et~al.}(2013)\citenamefont
  {Stranks}, \citenamefont {Eperon}, \citenamefont {Grancini}, \citenamefont
  {Menelaou}, \citenamefont {Alcocer}, \citenamefont {Leijtens}, \citenamefont
  {Herz}, \citenamefont {Petrozza},\ and\ \citenamefont
  {Snaith}}]{Stranks2013}%
  \BibitemOpen
  \bibfield  {author} {\bibinfo {author} {\bibfnamefont {S.~D.}\ \bibnamefont
  {Stranks}}, \bibinfo {author} {\bibfnamefont {G.~E.}\ \bibnamefont {Eperon}},
  \bibinfo {author} {\bibfnamefont {G.}~\bibnamefont {Grancini}}, \bibinfo
  {author} {\bibfnamefont {C.}~\bibnamefont {Menelaou}}, \bibinfo {author}
  {\bibfnamefont {M.~J.~P.}\ \bibnamefont {Alcocer}}, \bibinfo {author}
  {\bibfnamefont {T.}~\bibnamefont {Leijtens}}, \bibinfo {author}
  {\bibfnamefont {L.~M.}\ \bibnamefont {Herz}}, \bibinfo {author}
  {\bibfnamefont {A.}~\bibnamefont {Petrozza}}, \ and\ \bibinfo {author}
  {\bibfnamefont {H.~J.}\ \bibnamefont {Snaith}},\ }\href {\doibase
  10.1126/science.1243982} {\bibfield  {journal} {\bibinfo  {journal}
  {Science}\ }\textbf {\bibinfo {volume} {342}},\ \bibinfo {pages} {341}
  (\bibinfo {year} {2013})}\BibitemShut {NoStop}%
\bibitem [{\citenamefont {De~Wolf}\ \emph {et~al.}(2014)\citenamefont
  {De~Wolf}, \citenamefont {Holovsky}, \citenamefont {Moon}, \citenamefont
  {Löper}, \citenamefont {Niesen}, \citenamefont {Ledinsky}, \citenamefont
  {Haug}, \citenamefont {Yum},\ and\ \citenamefont {Ballif}}]{DeWolf2014}%
  \BibitemOpen
  \bibfield  {author} {\bibinfo {author} {\bibfnamefont {S.}~\bibnamefont
  {De~Wolf}}, \bibinfo {author} {\bibfnamefont {J.}~\bibnamefont {Holovsky}},
  \bibinfo {author} {\bibfnamefont {S.-J.}\ \bibnamefont {Moon}}, \bibinfo
  {author} {\bibfnamefont {P.}~\bibnamefont {Löper}}, \bibinfo {author}
  {\bibfnamefont {B.}~\bibnamefont {Niesen}}, \bibinfo {author} {\bibfnamefont
  {M.}~\bibnamefont {Ledinsky}}, \bibinfo {author} {\bibfnamefont {F.-J.}\
  \bibnamefont {Haug}}, \bibinfo {author} {\bibfnamefont {J.-H.}\ \bibnamefont
  {Yum}}, \ and\ \bibinfo {author} {\bibfnamefont {C.}~\bibnamefont {Ballif}},\
  }\href {\doibase 10.1021/jz500279b} {\bibfield  {journal} {\bibinfo
  {journal} {J. Phys. Chem. Lett.}\ }\textbf {\bibinfo {volume} {5}},\ \bibinfo
  {pages} {1035} (\bibinfo {year} {2014})}\BibitemShut {NoStop}%
\bibitem [{\citenamefont {Qiu}\ \emph {et~al.}(2015)\citenamefont {Qiu},
  \citenamefont {Paetzold}, \citenamefont {Gehlhaar}, \citenamefont {Smirnov},
  \citenamefont {Boyen}, \citenamefont {Tait}, \citenamefont {Conings},
  \citenamefont {Zhang}, \citenamefont {Nielsen}, \citenamefont {McCulloch},
  \citenamefont {Froyen}, \citenamefont {Heremans},\ and\ \citenamefont
  {Cheyns}}]{QIU2015}%
  \BibitemOpen
  \bibfield  {author} {\bibinfo {author} {\bibfnamefont {W.}~\bibnamefont
  {Qiu}}, \bibinfo {author} {\bibfnamefont {U.~W.}\ \bibnamefont {Paetzold}},
  \bibinfo {author} {\bibfnamefont {R.}~\bibnamefont {Gehlhaar}}, \bibinfo
  {author} {\bibfnamefont {V.}~\bibnamefont {Smirnov}}, \bibinfo {author}
  {\bibfnamefont {H.-G.}\ \bibnamefont {Boyen}}, \bibinfo {author}
  {\bibfnamefont {J.~G.}\ \bibnamefont {Tait}}, \bibinfo {author}
  {\bibfnamefont {B.}~\bibnamefont {Conings}}, \bibinfo {author} {\bibfnamefont
  {W.}~\bibnamefont {Zhang}}, \bibinfo {author} {\bibfnamefont {C.~B.}\
  \bibnamefont {Nielsen}}, \bibinfo {author} {\bibfnamefont {I.}~\bibnamefont
  {McCulloch}}, \bibinfo {author} {\bibfnamefont {L.}~\bibnamefont {Froyen}},
  \bibinfo {author} {\bibfnamefont {P.}~\bibnamefont {Heremans}}, \ and\
  \bibinfo {author} {\bibfnamefont {D.}~\bibnamefont {Cheyns}},\ }\href
  {\doibase 10.1039/C5TA07515G} {\bibfield  {journal} {\bibinfo  {journal} {J.
  Mater. Chem. A}\ }\textbf {\bibinfo {volume} {3}},\ \bibinfo {pages} {22824}
  (\bibinfo {year} {2015})}\BibitemShut {NoStop}%
\bibitem [{\citenamefont {{Jeon Nam Joong}}\ \emph {et~al.}(2015)\citenamefont
  {{Jeon Nam Joong}}, \citenamefont {{Noh Jun Hong}}, \citenamefont {{Yang Woon
  Seok}}, \citenamefont {{Kim Young Chan}}, \citenamefont {{Ryu Seungchan}},
  \citenamefont {{Seo Jangwon}},\ and\ \citenamefont {{Seok Sang
  Il}}}]{Joong2015}%
  \BibitemOpen
  \bibfield  {author} {\bibinfo {author} {\bibnamefont {{Jeon Nam Joong}}},
  \bibinfo {author} {\bibnamefont {{Noh Jun Hong}}}, \bibinfo {author}
  {\bibnamefont {{Yang Woon Seok}}}, \bibinfo {author} {\bibnamefont {{Kim
  Young Chan}}}, \bibinfo {author} {\bibnamefont {{Ryu Seungchan}}}, \bibinfo
  {author} {\bibnamefont {{Seo Jangwon}}}, \ and\ \bibinfo {author}
  {\bibnamefont {{Seok Sang Il}}},\ }\href {\doibase
  http://dx.doi.org/10.1038/nature14133 10.1038/nature14133} {\bibfield
  {journal} {\bibinfo  {journal} {Nature}\ }\textbf {\bibinfo {volume} {517}},\
  \bibinfo {pages} {476} (\bibinfo {year} {2015})}\BibitemShut {NoStop}%
\bibitem [{\citenamefont {Yang}\ \emph {et~al.}(2015)\citenamefont {Yang},
  \citenamefont {Noh}, \citenamefont {Jeon}, \citenamefont {Kim}, \citenamefont
  {Ryu}, \citenamefont {Seo},\ and\ \citenamefont {Seok}}]{Yang2015}%
  \BibitemOpen
  \bibfield  {author} {\bibinfo {author} {\bibfnamefont {W.~S.}\ \bibnamefont
  {Yang}}, \bibinfo {author} {\bibfnamefont {J.~H.}\ \bibnamefont {Noh}},
  \bibinfo {author} {\bibfnamefont {N.~J.}\ \bibnamefont {Jeon}}, \bibinfo
  {author} {\bibfnamefont {Y.~C.}\ \bibnamefont {Kim}}, \bibinfo {author}
  {\bibfnamefont {S.}~\bibnamefont {Ryu}}, \bibinfo {author} {\bibfnamefont
  {J.}~\bibnamefont {Seo}}, \ and\ \bibinfo {author} {\bibfnamefont {S.~I.}\
  \bibnamefont {Seok}},\ }\href {\doibase 10.1126/science.aaa9272} {\bibfield
  {journal} {\bibinfo  {journal} {Science}\ }\textbf {\bibinfo {volume}
  {348}},\ \bibinfo {pages} {1234} (\bibinfo {year} {2015})}\BibitemShut
  {NoStop}%
\bibitem [{\citenamefont {Saliba}\ \emph {et~al.}(2016)\citenamefont {Saliba},
  \citenamefont {Matsui}, \citenamefont {Seo}, \citenamefont {Domanski},
  \citenamefont {Correa-Baena}, \citenamefont {Nazeeruddin}, \citenamefont
  {Zakeeruddin}, \citenamefont {Tress}, \citenamefont {Abate}, \citenamefont
  {Hagfeldt},\ and\ \citenamefont {Gratzel}}]{Saliba2016}%
  \BibitemOpen
  \bibfield  {author} {\bibinfo {author} {\bibfnamefont {M.}~\bibnamefont
  {Saliba}}, \bibinfo {author} {\bibfnamefont {T.}~\bibnamefont {Matsui}},
  \bibinfo {author} {\bibfnamefont {J.-Y.}\ \bibnamefont {Seo}}, \bibinfo
  {author} {\bibfnamefont {K.}~\bibnamefont {Domanski}}, \bibinfo {author}
  {\bibfnamefont {J.-P.}\ \bibnamefont {Correa-Baena}}, \bibinfo {author}
  {\bibfnamefont {M.~K.}\ \bibnamefont {Nazeeruddin}}, \bibinfo {author}
  {\bibfnamefont {S.~M.}\ \bibnamefont {Zakeeruddin}}, \bibinfo {author}
  {\bibfnamefont {W.}~\bibnamefont {Tress}}, \bibinfo {author} {\bibfnamefont
  {A.}~\bibnamefont {Abate}}, \bibinfo {author} {\bibfnamefont
  {A.}~\bibnamefont {Hagfeldt}}, \ and\ \bibinfo {author} {\bibfnamefont
  {M.}~\bibnamefont {Gratzel}},\ }\href {\doibase 10.1039/C5EE03874J}
  {\bibfield  {journal} {\bibinfo  {journal} {Energy Environ. Sci.}\ }\textbf
  {\bibinfo {volume} {9}},\ \bibinfo {pages} {1989} (\bibinfo {year}
  {2016})}\BibitemShut {NoStop}%
\bibitem [{\citenamefont {Peng}\ \emph {et~al.}(2017)\citenamefont {Peng},
  \citenamefont {Wu}, \citenamefont {Ye}, \citenamefont {Jacobs}, \citenamefont
  {Shen}, \citenamefont {Fu}, \citenamefont {Wan}, \citenamefont {Duong},
  \citenamefont {Wu}, \citenamefont {Barugkin}, \citenamefont {Nguyen},
  \citenamefont {Zhong}, \citenamefont {Li}, \citenamefont {Lu}, \citenamefont
  {Liu}, \citenamefont {Lockrey}, \citenamefont {Weber}, \citenamefont
  {Catchpole},\ and\ \citenamefont {White}}]{Peng2017}%
  \BibitemOpen
  \bibfield  {author} {\bibinfo {author} {\bibfnamefont {J.}~\bibnamefont
  {Peng}}, \bibinfo {author} {\bibfnamefont {Y.}~\bibnamefont {Wu}}, \bibinfo
  {author} {\bibfnamefont {W.}~\bibnamefont {Ye}}, \bibinfo {author}
  {\bibfnamefont {D.~A.}\ \bibnamefont {Jacobs}}, \bibinfo {author}
  {\bibfnamefont {H.}~\bibnamefont {Shen}}, \bibinfo {author} {\bibfnamefont
  {X.}~\bibnamefont {Fu}}, \bibinfo {author} {\bibfnamefont {Y.}~\bibnamefont
  {Wan}}, \bibinfo {author} {\bibfnamefont {T.}~\bibnamefont {Duong}}, \bibinfo
  {author} {\bibfnamefont {N.}~\bibnamefont {Wu}}, \bibinfo {author}
  {\bibfnamefont {C.}~\bibnamefont {Barugkin}}, \bibinfo {author}
  {\bibfnamefont {H.~T.}\ \bibnamefont {Nguyen}}, \bibinfo {author}
  {\bibfnamefont {D.}~\bibnamefont {Zhong}}, \bibinfo {author} {\bibfnamefont
  {J.}~\bibnamefont {Li}}, \bibinfo {author} {\bibfnamefont {T.}~\bibnamefont
  {Lu}}, \bibinfo {author} {\bibfnamefont {Y.}~\bibnamefont {Liu}}, \bibinfo
  {author} {\bibfnamefont {M.~N.}\ \bibnamefont {Lockrey}}, \bibinfo {author}
  {\bibfnamefont {K.~J.}\ \bibnamefont {Weber}}, \bibinfo {author}
  {\bibfnamefont {K.~R.}\ \bibnamefont {Catchpole}}, \ and\ \bibinfo {author}
  {\bibfnamefont {T.~P.}\ \bibnamefont {White}},\ }\href {\doibase
  10.1039/C7EE01096F} {\bibfield  {journal} {\bibinfo  {journal} {Energy
  Environ. Sci.}\ }\textbf {\bibinfo {volume} {10}},\ \bibinfo {pages} {1792}
  (\bibinfo {year} {2017})}\BibitemShut {NoStop}%
\bibitem [{\citenamefont {deQuilettes}\ \emph {et~al.}(2016)\citenamefont
  {deQuilettes}, \citenamefont {Zhang}, \citenamefont {Burlakov}, \citenamefont
  {Graham}, \citenamefont {Leijtens}, \citenamefont {Osherov}, \citenamefont
  {Bulovi{\'c}}, \citenamefont {Snaith}, \citenamefont {Ginger},\ and\
  \citenamefont {Stranks}}]{deQuilettes2016}%
  \BibitemOpen
  \bibfield  {author} {\bibinfo {author} {\bibfnamefont {D.~W.}\ \bibnamefont
  {deQuilettes}}, \bibinfo {author} {\bibfnamefont {W.}~\bibnamefont {Zhang}},
  \bibinfo {author} {\bibfnamefont {V.~M.}\ \bibnamefont {Burlakov}}, \bibinfo
  {author} {\bibfnamefont {D.~J.}\ \bibnamefont {Graham}}, \bibinfo {author}
  {\bibfnamefont {T.}~\bibnamefont {Leijtens}}, \bibinfo {author}
  {\bibfnamefont {A.}~\bibnamefont {Osherov}}, \bibinfo {author} {\bibfnamefont
  {V.}~\bibnamefont {Bulovi{\'c}}}, \bibinfo {author} {\bibfnamefont {H.~J.}\
  \bibnamefont {Snaith}}, \bibinfo {author} {\bibfnamefont {D.~S.}\
  \bibnamefont {Ginger}}, \ and\ \bibinfo {author} {\bibfnamefont {S.~D.}\
  \bibnamefont {Stranks}},\ }\href {\doibase
  http://dx.doi.org/10.1038/ncomms11683 10.1038/ncomms11683} {\bibfield
  {journal} {\bibinfo  {journal} {Nat. Commun.}\ }\textbf {\bibinfo {volume}
  {7}},\ \bibinfo {pages} {11683} (\bibinfo {year} {2016})}\BibitemShut
  {NoStop}%
\bibitem [{\citenamefont {Crothers}\ \emph {et~al.}(2017)\citenamefont
  {Crothers}, \citenamefont {Milot}, \citenamefont {Patel}, \citenamefont
  {Parrott}, \citenamefont {Schlipf}, \citenamefont {Müller-Buschbaum},
  \citenamefont {Johnston},\ and\ \citenamefont {Herz}}]{Crothers2017}%
  \BibitemOpen
  \bibfield  {author} {\bibinfo {author} {\bibfnamefont {T.~W.}\ \bibnamefont
  {Crothers}}, \bibinfo {author} {\bibfnamefont {R.~L.}\ \bibnamefont {Milot}},
  \bibinfo {author} {\bibfnamefont {J.~B.}\ \bibnamefont {Patel}}, \bibinfo
  {author} {\bibfnamefont {E.~S.}\ \bibnamefont {Parrott}}, \bibinfo {author}
  {\bibfnamefont {J.}~\bibnamefont {Schlipf}}, \bibinfo {author} {\bibfnamefont
  {P.}~\bibnamefont {Müller-Buschbaum}}, \bibinfo {author} {\bibfnamefont
  {M.~B.}\ \bibnamefont {Johnston}}, \ and\ \bibinfo {author} {\bibfnamefont
  {L.~M.}\ \bibnamefont {Herz}},\ }\href {\doibase
  10.1021/acs.nanolett.7b02834} {\bibfield  {journal} {\bibinfo  {journal}
  {Nano Lett.}\ }\textbf {\bibinfo {volume} {17}},\ \bibinfo {pages} {5782}
  (\bibinfo {year} {2017})}\BibitemShut {NoStop}%
\bibitem [{\citenamefont {Davies}\ \emph {et~al.}(2018)\citenamefont {Davies},
  \citenamefont {Filip}, \citenamefont {Patel}, \citenamefont {Crothers},
  \citenamefont {Verdi}, \citenamefont {Wright}, \citenamefont {Milot},
  \citenamefont {Giustino}, \citenamefont {Johnston},\ and\ \citenamefont
  {Herz}}]{Davies2018}%
  \BibitemOpen
  \bibfield  {author} {\bibinfo {author} {\bibfnamefont {C.~L.}\ \bibnamefont
  {Davies}}, \bibinfo {author} {\bibfnamefont {M.~R.}\ \bibnamefont {Filip}},
  \bibinfo {author} {\bibfnamefont {J.~B.}\ \bibnamefont {Patel}}, \bibinfo
  {author} {\bibfnamefont {T.~W.}\ \bibnamefont {Crothers}}, \bibinfo {author}
  {\bibfnamefont {C.}~\bibnamefont {Verdi}}, \bibinfo {author} {\bibfnamefont
  {A.~D.}\ \bibnamefont {Wright}}, \bibinfo {author} {\bibfnamefont {R.~L.}\
  \bibnamefont {Milot}}, \bibinfo {author} {\bibfnamefont {F.}~\bibnamefont
  {Giustino}}, \bibinfo {author} {\bibfnamefont {M.~B.}\ \bibnamefont
  {Johnston}}, \ and\ \bibinfo {author} {\bibfnamefont {L.~M.}\ \bibnamefont
  {Herz}},\ }\href {\doibase 10.1038/s41467-017-02670-2} {\bibfield  {journal}
  {\bibinfo  {journal} {Nat. Commun.}\ }\textbf {\bibinfo {volume} {9}},\
  \bibinfo {pages} {293} (\bibinfo {year} {2018})}\BibitemShut {NoStop}%
\bibitem [{\citenamefont {Richter}\ \emph {et~al.}(2016)\citenamefont
  {Richter}, \citenamefont {Abdi-Jalebi}, \citenamefont {Sadhanala},
  \citenamefont {Tabachnyk}, \citenamefont {Rivett}, \citenamefont
  {Pazos-Out\'{o}n}, \citenamefont {G\"{o}del}, \citenamefont {Price},
  \citenamefont {Deschler},\ and\ \citenamefont {Friend}}]{Richter2016}%
  \BibitemOpen
  \bibfield  {author} {\bibinfo {author} {\bibfnamefont {J.~M.}\ \bibnamefont
  {Richter}}, \bibinfo {author} {\bibfnamefont {M.}~\bibnamefont
  {Abdi-Jalebi}}, \bibinfo {author} {\bibfnamefont {A.}~\bibnamefont
  {Sadhanala}}, \bibinfo {author} {\bibfnamefont {M.}~\bibnamefont
  {Tabachnyk}}, \bibinfo {author} {\bibfnamefont {J.~P.~H.}\ \bibnamefont
  {Rivett}}, \bibinfo {author} {\bibfnamefont {L.~M.}\ \bibnamefont
  {Pazos-Out\'{o}n}}, \bibinfo {author} {\bibfnamefont {K.~C.}\ \bibnamefont
  {G\"{o}del}}, \bibinfo {author} {\bibfnamefont {M.}~\bibnamefont {Price}},
  \bibinfo {author} {\bibfnamefont {F.}~\bibnamefont {Deschler}}, \ and\
  \bibinfo {author} {\bibfnamefont {R.~H.}\ \bibnamefont {Friend}},\ }\href
  {http://dx.doi.org/10.1038/ncomms13941} {\bibfield  {journal} {\bibinfo
  {journal} {Nat. Commun.}\ }\textbf {\bibinfo {volume} {7}},\ \bibinfo {pages}
  {13941 EP } (\bibinfo {year} {2016})}\BibitemShut {NoStop}%
\bibitem [{\citenamefont {Pazos-Out{\'o}n}\ \emph {et~al.}(2016)\citenamefont
  {Pazos-Out{\'o}n}, \citenamefont {Szumilo}, \citenamefont {Lamboll},
  \citenamefont {Richter}, \citenamefont {Crespo-Quesada}, \citenamefont
  {Abdi-Jalebi}, \citenamefont {Beeson}, \citenamefont {Vru{\'c}ini{\'c}},
  \citenamefont {Alsari}, \citenamefont {Snaith}, \citenamefont {Ehrler},
  \citenamefont {Friend},\ and\ \citenamefont {Deschler}}]{Pazos2016}%
  \BibitemOpen
  \bibfield  {author} {\bibinfo {author} {\bibfnamefont {L.~M.}\ \bibnamefont
  {Pazos-Out{\'o}n}}, \bibinfo {author} {\bibfnamefont {M.}~\bibnamefont
  {Szumilo}}, \bibinfo {author} {\bibfnamefont {R.}~\bibnamefont {Lamboll}},
  \bibinfo {author} {\bibfnamefont {J.~M.}\ \bibnamefont {Richter}}, \bibinfo
  {author} {\bibfnamefont {M.}~\bibnamefont {Crespo-Quesada}}, \bibinfo
  {author} {\bibfnamefont {M.}~\bibnamefont {Abdi-Jalebi}}, \bibinfo {author}
  {\bibfnamefont {H.~J.}\ \bibnamefont {Beeson}}, \bibinfo {author}
  {\bibfnamefont {M.}~\bibnamefont {Vru{\'c}ini{\'c}}}, \bibinfo {author}
  {\bibfnamefont {M.}~\bibnamefont {Alsari}}, \bibinfo {author} {\bibfnamefont
  {H.~J.}\ \bibnamefont {Snaith}}, \bibinfo {author} {\bibfnamefont
  {B.}~\bibnamefont {Ehrler}}, \bibinfo {author} {\bibfnamefont {R.~H.}\
  \bibnamefont {Friend}}, \ and\ \bibinfo {author} {\bibfnamefont
  {F.}~\bibnamefont {Deschler}},\ }\href {\doibase 10.1126/science.aaf1168}
  {\bibfield  {journal} {\bibinfo  {journal} {Science}\ }\textbf {\bibinfo
  {volume} {351}},\ \bibinfo {pages} {1430} (\bibinfo {year}
  {2016})}\BibitemShut {NoStop}%
\bibitem [{\citenamefont {Tress}(2017)}]{Tress2017}%
  \BibitemOpen
  \bibfield  {author} {\bibinfo {author} {\bibfnamefont {W.}~\bibnamefont
  {Tress}},\ }\href {\doibase 10.1002/aenm.201602358} {\bibfield  {journal}
  {\bibinfo  {journal} {Adv. Energy Mater.}\ }\textbf {\bibinfo {volume} {7}},\
  \bibinfo {pages} {1602358} (\bibinfo {year} {2017})}\BibitemShut {NoStop}%
\bibitem [{\citenamefont {Kirchartz}\ and\ \citenamefont
  {Rau}(2018)}]{Kirchartz2018}%
  \BibitemOpen
  \bibfield  {author} {\bibinfo {author} {\bibfnamefont {T.}~\bibnamefont
  {Kirchartz}}\ and\ \bibinfo {author} {\bibfnamefont {U.}~\bibnamefont
  {Rau}},\ }\href@noop {} {\bibfield  {journal} {\bibinfo  {journal} {Adv.
  Energy Mater.}\ }\textbf {\bibinfo {volume} {0}},\ \bibinfo {pages} {1703385}
  (\bibinfo {year} {2018})}\BibitemShut {NoStop}%
\bibitem [{\citenamefont {Shockley}\ and\ \citenamefont
  {Queisser}(1961)}]{SQ1961}%
  \BibitemOpen
  \bibfield  {author} {\bibinfo {author} {\bibfnamefont {W.}~\bibnamefont
  {Shockley}}\ and\ \bibinfo {author} {\bibfnamefont {H.~J.}\ \bibnamefont
  {Queisser}},\ }\href {\doibase 10.1063/1.1736034} {\bibfield  {journal}
  {\bibinfo  {journal} {Journal of Applied Physics}\ }\textbf {\bibinfo
  {volume} {32}},\ \bibinfo {pages} {510} (\bibinfo {year} {1961})}\BibitemShut
  {NoStop}%
\bibitem [{\citenamefont {W{\"u}rfel}(2007)}]{Wurfel2007}%
  \BibitemOpen
  \bibfield  {author} {\bibinfo {author} {\bibfnamefont {P.}~\bibnamefont
  {W{\"u}rfel}},\ }\href {\doibase 10.1002/9783527618545.ch2} {\emph {\bibinfo
  {title} {{Physics of Solar Cells}}}}\ (\bibinfo  {publisher} {Wiley-VCH
  Verlag GmbH},\ \bibinfo {year} {2007})\ pp.\ \bibinfo {pages}
  {9--35}\BibitemShut {NoStop}%
\bibitem [{\citenamefont {Kirchhoff}(1860)}]{Kirchoff1860}%
  \BibitemOpen
  \bibfield  {author} {\bibinfo {author} {\bibfnamefont {G.}~\bibnamefont
  {Kirchhoff}},\ }\href {\doibase 10.1002/andp.18601850205} {\bibfield
  {journal} {\bibinfo  {journal} {Annalen der Physik}\ }\textbf {\bibinfo
  {volume} {185}},\ \bibinfo {pages} {275} (\bibinfo {year}
  {1860})}\BibitemShut {NoStop}%
\bibitem [{\citenamefont {Tiedje}\ \emph {et~al.}(1984)\citenamefont {Tiedje},
  \citenamefont {Yablonovitch}, \citenamefont {Cody},\ and\ \citenamefont
  {Brooks}}]{Tiedje1984}%
  \BibitemOpen
  \bibfield  {author} {\bibinfo {author} {\bibfnamefont {T.}~\bibnamefont
  {Tiedje}}, \bibinfo {author} {\bibfnamefont {E.}~\bibnamefont
  {Yablonovitch}}, \bibinfo {author} {\bibfnamefont {G.~D.}\ \bibnamefont
  {Cody}}, \ and\ \bibinfo {author} {\bibfnamefont {B.~G.}\ \bibnamefont
  {Brooks}},\ }\href {\doibase 10.1109/T-ED.1984.21594} {\bibfield  {journal}
  {\bibinfo  {journal} {IEEE Trans. Electron Devices}\ }\textbf {\bibinfo
  {volume} {31}},\ \bibinfo {pages} {711} (\bibinfo {year} {1984})}\BibitemShut
  {NoStop}%
\bibitem [{\citenamefont {Kirchartz}\ and\ \citenamefont
  {Rau}(2008)}]{Kirchartz2008}%
  \BibitemOpen
  \bibfield  {author} {\bibinfo {author} {\bibfnamefont {T.}~\bibnamefont
  {Kirchartz}}\ and\ \bibinfo {author} {\bibfnamefont {U.}~\bibnamefont
  {Rau}},\ }\href {\doibase 10.1002/pssa.200880458} {\bibfield  {journal}
  {\bibinfo  {journal} {Phys. Status Solidi A}\ }\textbf {\bibinfo {volume}
  {205}},\ \bibinfo {pages} {2737} (\bibinfo {year} {2008})}\BibitemShut
  {NoStop}%
\bibitem [{\citenamefont {Rau}(2007)}]{Rau2007}%
  \BibitemOpen
  \bibfield  {author} {\bibinfo {author} {\bibfnamefont {U.}~\bibnamefont
  {Rau}},\ }\href {\doibase 10.1103/PhysRevB.76.085303} {\bibfield  {journal}
  {\bibinfo  {journal} {Phys. Rev. B}\ }\textbf {\bibinfo {volume} {76}},\
  \bibinfo {pages} {085303} (\bibinfo {year} {2007})}\BibitemShut {NoStop}%
\bibitem [{\citenamefont {Sze}\ and\ \citenamefont {Ng}(2006)}]{Sze2006}%
  \BibitemOpen
  \bibfield  {author} {\bibinfo {author} {\bibfnamefont {S.}~\bibnamefont
  {Sze}}\ and\ \bibinfo {author} {\bibfnamefont {K.~K.}\ \bibnamefont {Ng}},\
  }\href {\doibase 10.1002/9780470068328.ch1} {\emph {\bibinfo {title}
  {{Physics of Semiconductor Devices}}}}\ (\bibinfo  {publisher} {John Wiley \&
  Sons, Inc.},\ \bibinfo {year} {2006})\ pp.\ \bibinfo {pages}
  {5--75}\BibitemShut {NoStop}%
\bibitem [{\citenamefont {{Green}}(1982)}]{Green1982B}%
  \BibitemOpen
  \bibfield  {author} {\bibinfo {author} {\bibfnamefont {M.~A.}\ \bibnamefont
  {{Green}}},\ }\href@noop {} {\emph {\bibinfo {title} {{Solar cells: Operating
  principles, technology, and system applications}}}}\ (\bibinfo  {publisher}
  {Prentice-Hall, Inc.},\ \bibinfo {year} {1982})\BibitemShut {NoStop}%
\bibitem [{\citenamefont {Paulus}\ \emph {et~al.}(2000)\citenamefont {Paulus},
  \citenamefont {Gay-Balmaz},\ and\ \citenamefont {Martin}}]{Paulus2000}%
  \BibitemOpen
  \bibfield  {author} {\bibinfo {author} {\bibfnamefont {M.}~\bibnamefont
  {Paulus}}, \bibinfo {author} {\bibfnamefont {P.}~\bibnamefont {Gay-Balmaz}},
  \ and\ \bibinfo {author} {\bibfnamefont {O.~J.~F.}\ \bibnamefont {Martin}},\
  }\href {\doibase 10.1103/PhysRevE.62.5797} {\bibfield  {journal} {\bibinfo
  {journal} {Phys. Rev. E}\ }\textbf {\bibinfo {volume} {62}},\ \bibinfo
  {pages} {5797} (\bibinfo {year} {2000})}\BibitemShut {NoStop}%
\bibitem [{\citenamefont {Novotny}\ and\ \citenamefont
  {Hecht}(2006)}]{Novotny12Book}%
  \BibitemOpen
  \bibfield  {author} {\bibinfo {author} {\bibfnamefont {L.}~\bibnamefont
  {Novotny}}\ and\ \bibinfo {author} {\bibfnamefont {B.}~\bibnamefont
  {Hecht}},\ }\href@noop {} {\emph {\bibinfo {title} {Principles of
  nano-optics}}}\ (\bibinfo  {publisher} {Cambridge},\ \bibinfo {year} {2006})\
  p.\ \bibinfo {pages} {266}\BibitemShut {NoStop}%
\bibitem [{\citenamefont {Staub}\ \emph {et~al.}(2016)\citenamefont {Staub},
  \citenamefont {Hempel}, \citenamefont {Hebig}, \citenamefont {Mock},
  \citenamefont {Paetzold}, \citenamefont {Rau}, \citenamefont {Unold},\ and\
  \citenamefont {Kirchartz}}]{Staub2016}%
  \BibitemOpen
  \bibfield  {author} {\bibinfo {author} {\bibfnamefont {F.}~\bibnamefont
  {Staub}}, \bibinfo {author} {\bibfnamefont {H.}~\bibnamefont {Hempel}},
  \bibinfo {author} {\bibfnamefont {J.-C.}\ \bibnamefont {Hebig}}, \bibinfo
  {author} {\bibfnamefont {J.}~\bibnamefont {Mock}}, \bibinfo {author}
  {\bibfnamefont {U.~W.}\ \bibnamefont {Paetzold}}, \bibinfo {author}
  {\bibfnamefont {U.}~\bibnamefont {Rau}}, \bibinfo {author} {\bibfnamefont
  {T.}~\bibnamefont {Unold}}, \ and\ \bibinfo {author} {\bibfnamefont
  {T.}~\bibnamefont {Kirchartz}},\ }\href {\doibase
  10.1103/PhysRevApplied.6.044017} {\bibfield  {journal} {\bibinfo  {journal}
  {Phys. Rev. Appl.}\ }\textbf {\bibinfo {volume} {6}},\ \bibinfo {pages}
  {044017} (\bibinfo {year} {2016})}\BibitemShut {NoStop}%
\bibitem [{\citenamefont {Aeberhard}\ and\ \citenamefont
  {Rau}(2017)}]{Aeberhard2017}%
  \BibitemOpen
  \bibfield  {author} {\bibinfo {author} {\bibfnamefont {U.}~\bibnamefont
  {Aeberhard}}\ and\ \bibinfo {author} {\bibfnamefont {U.}~\bibnamefont
  {Rau}},\ }\href@noop {} {\bibfield  {journal} {\bibinfo  {journal} {Phys.
  Rev. Lett.}\ }\textbf {\bibinfo {volume} {118}},\ \bibinfo {pages} {247702}
  (\bibinfo {year} {2017})}\BibitemShut {NoStop}%
\bibitem [{\citenamefont {van Eerden}\ \emph {et~al.}(2017)\citenamefont {van
  Eerden}, \citenamefont {Jaysankar}, \citenamefont {Hadipour}, \citenamefont
  {Merckx}, \citenamefont {Schermer}, \citenamefont {Aernouts}, \citenamefont
  {Poortmans},\ and\ \citenamefont {Paetzold}}]{vanEerden2017}%
  \BibitemOpen
  \bibfield  {author} {\bibinfo {author} {\bibfnamefont {M.}~\bibnamefont {van
  Eerden}}, \bibinfo {author} {\bibfnamefont {M.}~\bibnamefont {Jaysankar}},
  \bibinfo {author} {\bibfnamefont {A.}~\bibnamefont {Hadipour}}, \bibinfo
  {author} {\bibfnamefont {T.}~\bibnamefont {Merckx}}, \bibinfo {author}
  {\bibfnamefont {J.~J.}\ \bibnamefont {Schermer}}, \bibinfo {author}
  {\bibfnamefont {T.}~\bibnamefont {Aernouts}}, \bibinfo {author}
  {\bibfnamefont {J.}~\bibnamefont {Poortmans}}, \ and\ \bibinfo {author}
  {\bibfnamefont {U.~W.}\ \bibnamefont {Paetzold}},\ }\href {\doibase
  10.1002/adom.201700151} {\bibfield  {journal} {\bibinfo  {journal} {Advanced
  Optical Materials}\ }\textbf {\bibinfo {volume} {5}},\ \bibinfo {pages}
  {1700151} (\bibinfo {year} {2017})}\BibitemShut {NoStop}%
\bibitem [{\citenamefont {Born}\ and\ \citenamefont
  {Wolf}(1980)}]{BORN1980665}%
  \BibitemOpen
  \bibfield  {author} {\bibinfo {author} {\bibfnamefont {M.}~\bibnamefont
  {Born}}\ and\ \bibinfo {author} {\bibfnamefont {E.}~\bibnamefont {Wolf}},\
  }\href {\doibase 10.1016/B978-0-08-026482-0.50021-9} {\emph {\bibinfo {title}
  {{Principles of Optics}}}},\ \bibinfo {edition} {sixth edition}\ ed.,\ edited
  by\ \bibinfo {editor} {\bibfnamefont {M.}~\bibnamefont {Born}}\ and\ \bibinfo
  {editor} {\bibfnamefont {E.}~\bibnamefont {Wolf}}\ (\bibinfo  {publisher}
  {Pergamon},\ \bibinfo {year} {1980})\BibitemShut {NoStop}%
\bibitem [{\citenamefont {{Tai}}\ and\ \citenamefont
  {{Collin}}(2000)}]{Tai2000}%
  \BibitemOpen
  \bibfield  {author} {\bibinfo {author} {\bibfnamefont {C.~T.}\ \bibnamefont
  {{Tai}}}\ and\ \bibinfo {author} {\bibfnamefont {R.~E.}\ \bibnamefont
  {{Collin}}},\ }\href {\doibase 10.1109/8.899665} {\bibfield  {journal}
  {\bibinfo  {journal} {IEEE Trans. Antennas Propag.}\ }\textbf {\bibinfo
  {volume} {48}},\ \bibinfo {pages} {1501} (\bibinfo {year}
  {2000})}\BibitemShut {NoStop}%
\bibitem [{\citenamefont {Barnett}\ \emph {et~al.}(1992)\citenamefont
  {Barnett}, \citenamefont {Huttner},\ and\ \citenamefont
  {Loudon}}]{Barnett1992}%
  \BibitemOpen
  \bibfield  {author} {\bibinfo {author} {\bibfnamefont {S.~M.}\ \bibnamefont
  {Barnett}}, \bibinfo {author} {\bibfnamefont {B.}~\bibnamefont {Huttner}}, \
  and\ \bibinfo {author} {\bibfnamefont {R.}~\bibnamefont {Loudon}},\ }\href
  {\doibase 10.1103/PhysRevLett.68.3698} {\bibfield  {journal} {\bibinfo
  {journal} {Phys. Rev. Lett.}\ }\textbf {\bibinfo {volume} {68}},\ \bibinfo
  {pages} {3698} (\bibinfo {year} {1992})}\BibitemShut {NoStop}%
\bibitem [{\citenamefont {Scheel}\ \emph {et~al.}(1999)\citenamefont {Scheel},
  \citenamefont {Kn\"oll},\ and\ \citenamefont {Welsch}}]{Schell1999}%
  \BibitemOpen
  \bibfield  {author} {\bibinfo {author} {\bibfnamefont {S.}~\bibnamefont
  {Scheel}}, \bibinfo {author} {\bibfnamefont {L.}~\bibnamefont {Kn\"oll}}, \
  and\ \bibinfo {author} {\bibfnamefont {D.-G.}\ \bibnamefont {Welsch}},\
  }\href {\doibase 10.1103/PhysRevA.60.4094} {\bibfield  {journal} {\bibinfo
  {journal} {Phys. Rev. A}\ }\textbf {\bibinfo {volume} {60}},\ \bibinfo
  {pages} {4094} (\bibinfo {year} {1999})}\BibitemShut {NoStop}%
\bibitem [{\citenamefont {Juzeliūnas}(2006)}]{Juzeliunas2006}%
  \BibitemOpen
  \bibfield  {author} {\bibinfo {author} {\bibfnamefont {G.}~\bibnamefont
  {Juzeliūnas}},\ }\href {http://stacks.iop.org/0953-4075/39/i=15/a=S10}
  {\bibfield  {journal} {\bibinfo  {journal} {J. Phys. B}\ }\textbf {\bibinfo
  {volume} {39}},\ \bibinfo {pages} {S627} (\bibinfo {year}
  {2006})}\BibitemShut {NoStop}%
\bibitem [{\citenamefont {Rikken}\ and\ \citenamefont
  {Kessener}(1995)}]{Rikken1995}%
  \BibitemOpen
  \bibfield  {author} {\bibinfo {author} {\bibfnamefont {G.~L. J.~A.}\
  \bibnamefont {Rikken}}\ and\ \bibinfo {author} {\bibfnamefont {Y.~A. R.~R.}\
  \bibnamefont {Kessener}},\ }\href {\doibase 10.1103/PhysRevLett.74.880}
  {\bibfield  {journal} {\bibinfo  {journal} {Phys. Rev. Lett.}\ }\textbf
  {\bibinfo {volume} {74}},\ \bibinfo {pages} {880} (\bibinfo {year}
  {1995})}\BibitemShut {NoStop}%
\bibitem [{\citenamefont {Schuurmans}\ \emph {et~al.}(1998)\citenamefont
  {Schuurmans}, \citenamefont {de~Lang}, \citenamefont {Wegdam}, \citenamefont
  {Sprik},\ and\ \citenamefont {Lagendijk}}]{Schuurmans1998}%
  \BibitemOpen
  \bibfield  {author} {\bibinfo {author} {\bibfnamefont {F.~J.~P.}\
  \bibnamefont {Schuurmans}}, \bibinfo {author} {\bibfnamefont {D.~T.~N.}\
  \bibnamefont {de~Lang}}, \bibinfo {author} {\bibfnamefont {G.~H.}\
  \bibnamefont {Wegdam}}, \bibinfo {author} {\bibfnamefont {R.}~\bibnamefont
  {Sprik}}, \ and\ \bibinfo {author} {\bibfnamefont {A.}~\bibnamefont
  {Lagendijk}},\ }\href {\doibase 10.1103/PhysRevLett.80.5077} {\bibfield
  {journal} {\bibinfo  {journal} {Phys. Rev. Lett.}\ }\textbf {\bibinfo
  {volume} {80}},\ \bibinfo {pages} {5077} (\bibinfo {year}
  {1998})}\BibitemShut {NoStop}%
\end{thebibliography}%

\end{document}